\documentclass[11pt]{article}

\usepackage[T1]{fontenc}
\usepackage[utf8]{inputenc}
\usepackage{amsmath,amssymb,amsthm}
\usepackage{hyperref}
\usepackage{graphicx}
\usepackage{booktabs}
\usepackage[a4paper,margin=3cm]{geometry}

\title{Inferring cellular heterogeneity with mixture models for DNA methylation rates}
\author{
    Hugo Barbot\\
    IRMAR - UMR CNRS 6625\\
    \texttt{hugo.barbot@institut-agro.fr}
    \and
    Yuna Blum\\
    IGDR - UMR CNRS 6290\\
    $\qquad \quad$\texttt{yuna.blum@univ-rennes.fr}$\qquad \quad$
    \and
    Magali Richard\\
    LIG - UMR CNRS 5217\\
    \texttt{magali.richard@univ-grenoble-alpes.fr}
    \and
    David Causeur\\
    IRMAR - UMR CNRS 6625\\
    \texttt{david.causeur@institut-agro.fr}
}
\date{\today}

\begin{document}

\maketitle

\begin{abstract}
\noindent Cellular heterogeneity is a hallmark of biological tissues and plays a central role in disease progression, diagnosis, and prognosis. Yet, accurately characterizing this heterogeneity from bulk molecular profiles remains challenging because observed signals arise from mixtures of multiple cell populations. Cell deconvolution aim to recover the relative abundance of constituent cell types from such heterogeneous measurements, but most existing approaches implicitly rely on restrictive assumptions on residual errors, including independence, homoscedasticity, and normality. These assumptions are rarely satisfied in omics data, which are inherently bounded and overdispersed.
  In this work, we show that whole-genome cell-type specific DNA methylation profiles exhibit latent group structures that can substantially impair deconvolution accuracy when ignored. We therefore propose a mixture of non-negative Beta regression models estimated through an Expectation-Maximization algorithm for DNA methylation rates. Our framework naturally incorporates a feature selection mechanism through mixture component identification, making component selection a critical step of the inference procedure. 
  We further propose a dedicated criterion for component selection and assess the performance of the approach through an extensive comparative study across several \textit{in vitro} benchmark datasets. Our results demonstrate that deconvolution accuracy is highly sensitive to latent component structure and show that explicitly modeling this heterogeneity yields substantial improvements over standard whole-genome deconvolution strategies. Altogether, this work establishes mixture modeling of DNA methylation data as a powerful new direction for robust and accurate cell deconvolution.
\end{abstract}


\section{Introduction}
\label{s:intro}

The basic principle of cell deconvolution is to infer the underlying cellular heterogeneity from molecular profiles of pooled cell populations, commonly referred to as bulk sample \cite{nguyen2024fourteen}. Among the various types of omic data used for this purpose, transcriptomic profiles obtained through RNA sequencing and DNA methylation levels are the most prevalent \cite{hannon2024transcriptome,amblard2025robust}. The cell proportions estimated within a bulk sample by a deconvolution model reflect progression of disease state and are therefore useful for improved diagnostic and prognosis.

Cell deconvolution can be performed without cell type reference profiles in an unsupervised manner, or with well-characterized molecular reference profiles for a known set of cell populations in a supervised manner. This work focuses exclusively on supervised approaches applied to DNA methylation data, one of the two most frequently used omics data types for deconvolution \cite{hannon2024transcriptome,garmire2024challenges}.

Whatever the omic data type used for cell deconvolution, standard supervised models assume that the bulk profile results from a linear combination of the reference cell-type profiles, where the unknown mixing proportions represent the relative abundance of each cell type in the bulk sample, up to an additive error term. As a result of the linear combination hypothesis, most statistical methods used for cellular deconvolution are based on extensions of the Ordinary Least Squares (OLS) optimization algorithm \cite{garmire2024challenges,avila2020benchmarking,decamps2021deconbench}. Whereas unconstrained least-squares optimization is well-studied and guarantees desirable properties, such as unbiasedness and minimum variance under the standard assumptions of the linear regression model, this optimization problem is substantially more challenging in the present cell deconvolution issue due to the non-negativity and sum-to-one constraints on mixing coefficients. This explains the large diversity of algorithms currently available for cell deconvolution. 

A confusing point in cell deconvolution methods is that genes or CpG probes are considered as statistical units. Whereas in standard approaches of omic data analysis, genes or CpG probes are usually considered as features measured on independent statistical units being different biological samples. The use of OLS-based algorithms for cell deconvolution implicitly relies on the standard assumptions of linear regression models, among which independence across those statistical units, homoscedasticity and normality of residual errors are particularly important. However, in the context of DNA methylation data, these assumptions are highly questionable: methylation rates are indeed bounded in the interval $[0,1]$, notoriously exhibiting pronounced heteroscedasticity and shaped by local chromatin context and regulatory mechanisms, leading to a structured dependence pattern along the genome.

To address the interval restriction of DNA methylation measurements to $[0,1]$ and the presence of heteroscedasticity, we propose a model in which DNA methylation rates follow a Beta distribution conditional on the methylation profiles of the reference cell populations. Modeling DNA methylation data using the Beta distribution has been shown to improve accuracy in a range of statistical genomics applications, including differential methylation analysis \cite{majumdar2024_Rau_Beta_for_DNA,triche2016beta}. Beyond single-component models, mixtures of Beta distributions have been proposed to address heterogeneity in methylation levels across the genome, with applications to dimensionality reduction and classification \cite{laurila2011betamixt_dim_reduc}, as well as to integrative analyses of DNA methylation and gene expression \cite{gevaert2015methylmix}. In these studies, Beta mixture models capture distinct methylation modes, typically corresponding to hypomethylated and hypermethylated states, and in some cases an additional intermediate class representing moderately methylated loci.

Our approach is further designed to capture heterogeneous signals across the genome by means of a mixture of non-negative Beta regression models. In contrast to existing Beta mixture approaches, the latent components in our model are not designed to represent distinct modes of DNA methylation levels. Instead, they induce a partition of the genome into subsets of features that differ in their ability to accurately estimate the cellular composition of bulk samples. Under this perspective, the mixture formulation provides a general and principled framework for bulk-specific feature selection (where "features" refer here to statistical units, i.e., genes or CpG sites), a problem that has been identified as a key challenge in cell deconvolution methodologies \cite{garmire2024challenges}.

Most existing feature selection strategies are developed for collections of bulk samples sharing a common reference matrix and are applied as a preprocessing step prior to proportion estimation, independently of the underlying deconvolution model and its associated inference procedure. By contrast, our bulk-specific feature selection approach is intrinsically embedded within the deconvolution model itself. This formulation naturally raises the question of how to identify the mixture component that yields the most accurate estimation of cell type proportions. To address this issue, we propose and develop a statistical criterion specifically tailored to compare latent components of the non-negative Beta regression mixture model in terms of their estimation accuracy.

Section \ref{s:background} introduces the basic principles of non-negative regression models and an overview of the associated estimation algorithms used for cell deconvolution. It also critically examines the assumptions underlying these methods, using as an illustration a benchmark dataset generated for the study of pancreatic ductal adenocarcinoma (PDAC) \cite{amblard2025robust}. In Section \ref{s:Mixt-NNBR}, we introduce a greedy coordinate descent (GCD) algorithm for maximum-likelihood estimation in the non-negative Beta regression (NNBR) cell deconvolution model, together with a closed-form estimator of the precision parameter derived from an approximation of the estimating equations under large-precision asymptotics. Also, Expectation-Maximisation inference for mixtures of non-negative Beta regression models (M-NNBR) is presented in Section \ref{s:Mixt-NNBR}, whereas its associated criterion for feature selection is introduced and discussed in Section \ref{s:gene_selection}.

In Section \ref{s:comp_stud}, we evaluate the proposed deconvolution framework through a comparative study using the illustrative PDAC dataset as well as additional benchmark datasets. The paper concludes with a summary of the main findings and a discussion of perspectives for future work.

\section{Cell deconvolution using DNA methylation rates}
\label{s:background}

Bulk DNA methylation data from a recent study on Pancreatic Ductal Adenocarcinoma (PDAC), see \cite{amblard2025robust}, will be used throughout to illustrate the application of standard cell deconvolution algorithms and to critically examine their underlying assumptions. This dataset involves $n=30$ pooled samples composed of $p=9$ distinct cell populations with strictly \textit{in vitro} controlled proportions. This benchmark dataset consists of DNA methylation measurements at over 800,000 CpG sites across the genome, available both for each bulk sample and for the corresponding reference cell populations. To reduce the volume of data and enhance interpretability, DNA methylation rates are aggregated at the gene level by averaging the values of CpG sites located in the promoter region of each gene. As a result, gene-level methylation profiles covering $m=18,735$ features are available for each of the $n$ bulk samples. The reference signature matrix includes the DNA methylation rates of these $m$ genes across the $p$ reference pure cell populations. 
In the following, all bulks are considered and treated independently.

For $j=1,\ldots,m$, let $y_{j}$ denote the bulk DNA methylation rate of gene $j$, and define the bulk methylation profile as the vector $y=(y_{1}, \ldots, y_{m})'$. Similarly, let $x_{j}=(x_{j1},\ldots,x_{jp})', \ j = 1, \ldots, m$ be the vector of DNA methylation rates observed in the $p$ reference pure cell types. 

\begin{figure}[!ht]
  \begin{center}
  \includegraphics[width=0.6\textwidth]{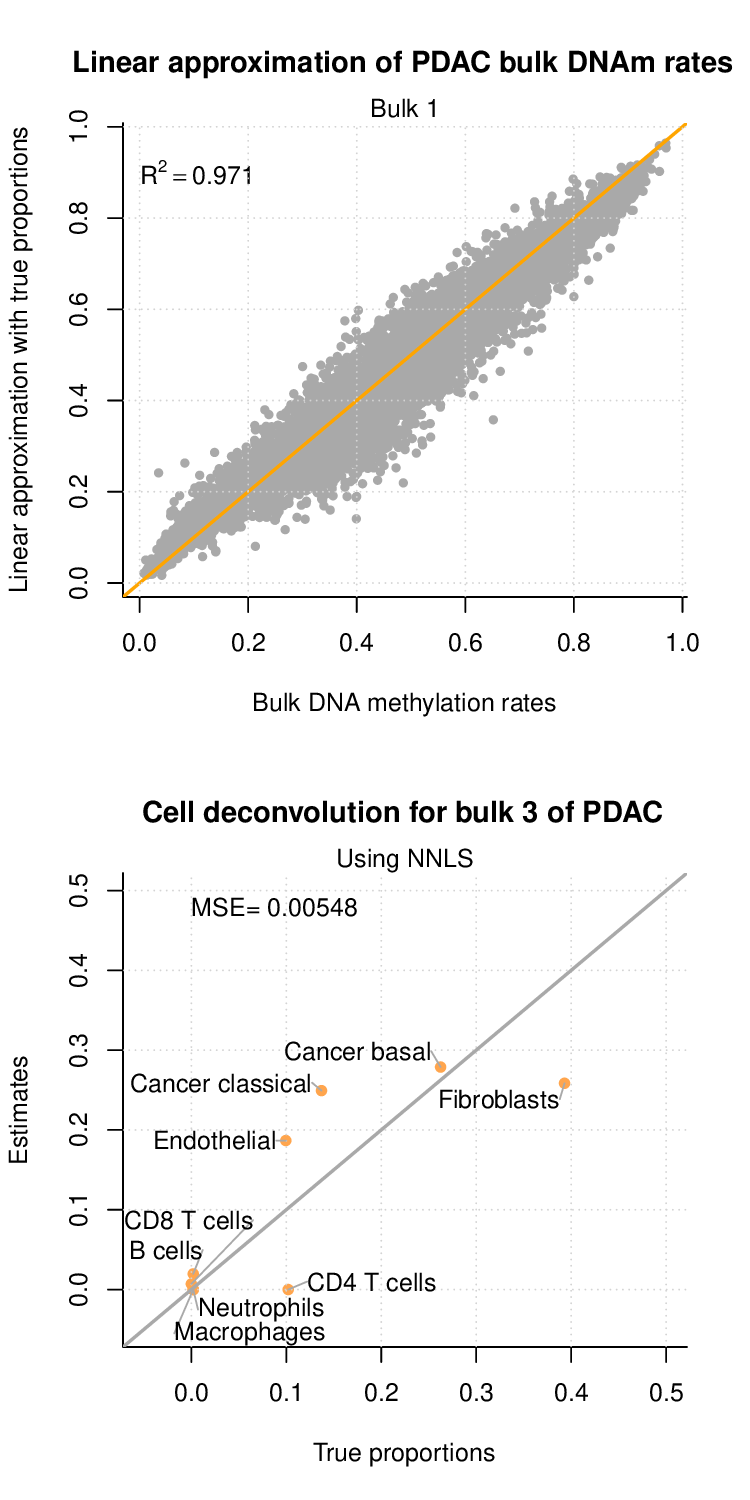}
  \end{center}
  \caption{PDAC study. Top plot: whole-genome Bulk 1 DNA methylation rates $y_{j}$ along with the linear combinations $\tilde{y}_{j}$ of the DNA methylation rates in the reference cell populations using the true proportions given in supplementary Table \ref{tab:true_beta_bulk_1}. The squared correlation between observed and approximated values ($R^{2}=0.971$) is indicated in the upper left corner of the plot. Bottom plot: non-negative least squares (NNLS) estimation of the cell type proportions for bulk 3 using \texttt{R} package \texttt{nnls} with one of the closest MSE to the median.}
  \label{fig:approx_bulk_1}
\end{figure}

Cell deconvolution methods assume that the bulk DNA methylation rates result from a weighted sum of the cell type specific DNA methylation rates, the weights being the proportions of the corresponding cell populations. For illustration of the validity of this linearity assumption over the genome, the top plot in Figure \ref{fig:approx_bulk_1} confirms the genomewide close approximation of the bulk DNA methylation rates $y$ in Bulk 1 of the PDAC study by the linear approximation $\tilde{y}_{j}$ of the DNA methylation rates in the nine reference cell populations using the true proportions given in supplementary table \ref{tab:true_beta_bulk_1}. Bulk 1 is chosen arbitrarily and similar plots can be produced for all bulks in the PDAC study. The mean of squared correlations between observed and linearly approximated values is 0.973 over the 30 bulks in the PDAC study.

Under the linearity assumption, most cell deconvolution algorithms are formulated as variants of the following constrained least-squares optimization problem:
\begin{eqnarray*}
  (\hat{\beta}_{0},\hat{\beta}') & = & \text{argmin}_{(\beta_{0},\beta')} \sum_{j=1}^{m} ( y_{j} - \beta_{0} - x'_{j} \beta )^{2} ,
\end{eqnarray*}
where $\beta_{0}$ is an intercept term, and $\beta$ is constrained to lie in the $p$-dimensional simplex ${\cal S}_{p} = \left\{ \beta = (\beta_{1}, \ldots , \beta_{p}), \ 0 \leq \beta_{r} \leq 1 , \sum_{r=1}^{p} \beta_{r} = 1 \right\}$. Non-negative least squares (NNLS) is the most used supervised deconvolution baseline in benchmark paper. NNLS estimations are performed via the \texttt{nnls} package in \texttt{R} \cite{nnls}, with posterior normalization to ensure estimated proportions sum to one. In the present situation, true proportions of cell types in each bulk are controlled and therefore can be assumed to be known. Then, cell deconvolution performance is evaluated using Mean Squared Error (MSE) of estimation, over the nine cell types.

Despite the close fit between observed and approximated bulk DNA methylation rates, the non-negative least squares (NNLS) estimates of cell type proportions remain inaccurate (bottom plot of figure \ref{fig:approx_bulk_1} and supplementary table \ref{tab:true_beta_bulk_1}). Across the 30 bulk samples of the \texttt{PDAC} dataset, NNLS achieves a median MSE of 0.00542 (range: 0.00082 to 0.01776). This is illustrated by bulk 3 in the PDAC study, for which MSE is one of the closest to the median MSE, fibroblast and CD4 T cell proportions are severely underestimated (25.85\% vs. a true value of 39.30\% and 0.00\% vs. 10.20\%, respectively), while endothelial cell and classical cancer cell proportions are substantially overestimated (18.67\% vs. a true value of 9.95\% and 24.93\% vs. 13.70\%, respectively). These discrepancies highlight the limitations of NNLS even in controlled benchmark settings where reference profiles are accurate and well characterized.

Optimality properties of unconstrained least-squares estimation are granted under the standard assumptions of linear regression models, among which homoscedasticity of the residuals. To illustrate the pronounced heteroscedasticity in DNA methylation data, we rely on the PDAC benchmark dataset introduced earlier and investigate the typical relationship between the conditional variance $v(x) = \text{Var}(Y \mid X = x)$ and the conditional mean $m(x) = \mathbb{E}(Y \mid X = x)$ of bulk DNA methylation rates, given the DNA methylation rates in the reference cell populations. Notably, the methylation profiles across reference cell types are highly correlated, with the first principal component $u(x)$ of the sample correlation matrix explaining 86.09\% of the total variance in the PDAC study. This allows for a simplified analysis in which the mean $m(x)$ and variance $v(x)$ can be approximated over this first principal component by $\tilde{m}(x) = \mathbb{E}(Y \mid u(X) = u(x))$ and $\tilde{v}(x) = \text{Var}(Y \mid u(X) = u(x))$, respectively.

Local estimates of $\tilde{m}(x)$ and $\tilde{v}(x)$ are obtained by computing sample means and variances of bulk DNA methylation rates over subsets of genes $j$ such that $u(x_j)$ lies within a local neighborhood of $u(x)$. The empirical relationship between conditional variances and means is displayed in the supplementary Figure \ref{fig:overdispersion_bulk_1} for Bulk 1 of the PDAC study chosen for an illustrative purpose, with results consistent across all bulks. The conditional variance $\tilde{v}(x)$ peaks when $\tilde{m}(x)$ is close to 0.5 and decreases markedly as $\tilde{m}(x)$ approaches 0 or 1, illustrating the heteroscedastic nature of the data.

Furthermore, many authors \cite{majumdar2024_Rau_Beta_for_DNA,triche2016beta} have suggested that the Beta distribution is more suited for modeling DNA methylation rates. In this context, we introduce a non-negative Beta regression model for cell deconvolution based on DNA methylation data.

\section{A mixture regression framework for DNA methylation deconvolution}
\label{s:Mixt-NNBR}

\subsection{Non-negative Beta regression (NNBR)}
\label{s:Mixt-NNBR1}

First, let us introduce the non-negative Beta regression (NNBR) model for cell deconvolution, with all bulks considered and treated independently. For gene $j = 1,\dots,m$, given the $p$-dimensional vectors $x_{j}=(x_{j1},\ldots,x_{jp})'$ of reference DNA methylation rates across the $p$ cell populations of interest, it is assumed that the bulk DNA methylation rates $y_{j}$ are distributed according to a Beta distribution with density \cite{triche2016beta,ferrari2004beta}:
  \begin{eqnarray}
    \varphi ( y_{j} \ | \ x_{j} ) & = & \frac{\Gamma(\phi)}{\Gamma (\mu_{j} \phi) \Gamma ( ( 1 - \mu_{j} ) \phi )} y_{j}^{\mu_{j} \phi - 1} 
    (1-y_{j})^{(1-\mu_{j})\phi - 1} , \label{eq:cellbetamod}
  \end{eqnarray}
where $\Gamma$ is the standard gamma function, $\phi>0$ is the precision parameter and $\mu_{j} = \mathbb{E} ( y_{j} \ | \ x_{j} ) = \beta_{0} + x'_{j} \beta$. Here, $\beta_{0}$ is an intercept parameter and $\beta=(\beta_{1},\ldots,\beta_{p})'$ is the vector of non-negative regression parameters.  

Model \eqref{eq:cellbetamod} corresponds to a particular case of the standard Beta regression model, in which the mean $\mu_{j}$ is typically linked to the linear predictor via a non-linear link function $g$ ensuring that $0 < \mu_{j} = g^{-1} (\beta_{0} + x'_{j} \beta) < 1$ \cite{ferrari2004beta,grun2012extended}. In contrast, our NNBR model requires the identity link function to directly models $\mu_{j}$ as a linear combination, as assumed in section \ref{s:background}. Therefore, $\mu_{j}$ is constrained to be positive and smaller than one, which requires a particular attention in the estimation procedure. Moreover, regression parameters $\beta_{r}, \ r = 1,\ldots,p,$ are assumed to be non-negative. 

Model \eqref{eq:cellbetamod} is intrinsically heteroscedastic, with the relationship between the conditional mean and variance controlled by the precision parameter $\phi$. Specifically, the conditional variance of the bulk DNA methylation rate $y_{j}$ given $x_{j}$ is given by:
  \begin{eqnarray}
    \text{Var} ( y_{j} \ | \ x_{j} ) & = & \frac{\mu_{j} ( 1 - \mu_{j} )}{1+\phi} .    \label{eq:varmean_betareg}
  \end{eqnarray}
When $\phi$ is close to zero, the conditional variance approaches that of a binary variable, reflecting large dispersion. Conversely, large values of $\phi$ induce underdispersion, resulting in more concentrated DNA methylation rate distributions. 

If $\theta=(\phi, \beta_{0}, \beta')$ denotes the vector of all parameters in model \eqref{eq:cellbetamod}, the log-likelihood is given by $\mathcal{L} ( \theta )=\sum_{j=1}^{m} \ell_{j} ( \mu_{j} , \phi )$, where:
  \begin{eqnarray} \label{eq:unweighted_LL}
    \ell_{j} ( \mu_{j} , \phi  ) & = & \text{log} \Gamma(\phi) - \text{log} \Gamma (\mu_{j} \phi) - \text{log}  \Gamma ( ( 1 - \mu_{j} ) \phi ) + \nonumber \\
  & & \bigl( \mu_{j} \phi - 1 ) \text{log} ( y_{j} ) + \bigl( ( 1 - \mu_{j} )  \phi - 1 ) \bigr) \text{log} ( 1 - y_{j} ) .
  \end{eqnarray}
    
A Greedy Coordinate Descent (GCD) algorithm is proposed to maximize the log-likelihood with respect to the precision parameter $\phi$ and the regression parameters $\beta_{0}$ and $\beta$, where the coordinates of $\beta$ are constrained to be non-negative (see Appendix \ref{app:gcd}). As in many cell deconvolution methods based on ordinary least squares (OLS), the sum-to-one constraint is enforced \textit{a posteriori} by normalizing the estimated proportion vector.

The GCD algorithm proceeds by iteratively updating one parameter at a time while holding the others fixed. This coordinate-wise strategy allows non-negativity to be enforced directly by setting any parameter to zero whenever its corresponding marginal maximization of the log-likelihood yields a negative value. Moreover, at each iteration, a rejection rule discards any gene $j$ for which the current fitted value of $\mu_{j}$ falls outside the open interval $(0,1)$, thereby ensuring compatibility with the Beta distribution assumption. In Appendix \ref{app:gcd}, the GCD algorithm is compared with a Fisher scoring alternative, enforcing non-negativity and the sum-to-one constraint \textit{a posteriori}, across all bulk samples from the four studies introduced in Section \ref{s:comp_stud}. This comparative analysis demonstrates the overall superiority of the GCD algorithm in terms of computation time and its ability to achieve maximum log-likelihood values that are comparable to or larger than those obtained via Fisher scoring (see supplementary Figure \ref{fig:app_gcd_versus_fs}).

By analogy with generalized linear models, the adequacy of the maximum-likelihood fit can be assessed using the deviance, ${\cal D} \;=\; 2 \bigl( {\cal L}_{\mathrm{sat}} - {\cal L}(\hat{\theta}) \bigr),$ where $\hat{\theta} = (\hat{\phi}, \hat{\beta}_{0}, \hat{\beta}')$ denotes the maximum-likelihood estimator of $\theta$, and ${\cal L}_{\mathrm{sat}}$ is the log-likelihood of the saturated model. If $\phi$ were known, the saturated fit for each gene $j$ would be obtained by the mean $\tilde{\mu}_{j}$ that maximizes its individual contribution $ \ell_{j}(\mu_{j},\phi)$. Following the recommendation of \cite{ferrari2004beta}, we approximate this quantity by the maximizer of $\ell_{j}(\mu_{j},\hat{\phi})$. In our implementation, the computation of $\tilde{\mu}_{j}$ relies on numerical maximization of $\ell_{j}(\mu_{j},\hat{\phi})$ using the \texttt{optimize} function in \texttt{R} \cite{brent2013algorithms}. As noted in \cite{ferrari2004beta}, when $\hat{\phi}$ is sufficiently large, the resulting maximizer $\tilde{\mu}_{j}$ is well approximated by the observed response $y_{j}$.

If model \eqref{eq:cellbetamod} were adequate at the genome-wide scale, the deviance residuals $ \hat{\varepsilon}_{j} \;=\; \operatorname{sign}(y_{j}-\hat{\mu}_{j})\,
  \sqrt{\,2\bigl(\ell_{j}(\tilde{\mu}_{j},\hat{\phi}) - \ell_{j}(\hat{\mu}_{j},\hat{\phi})\bigr)\,}$, with $\hat{\mu}_{j} = \hat{\beta}_{0} + x_{j}'\hat{\beta},$
would be expected to follow approximately a standard normal distribution. In practice, however, pronounced departures from normality are observed for all bulk samples in the studies introduced in Section \ref{s:comp_stud}.

To illustrate this point, supplementary figure \ref{fig:deviance_residuals} displays a histogram of the deviance residuals for Bulk 1 of the PDAC study introduced in Section \ref{s:background}. The pronounced departure from normality is representative of all bulks in this study. Notably, the empirical distribution is far better captured by a seven-component normal mixture model, with the number of components selected using the Bayesian Information Criterion (BIC).

\begin{figure}[!ht]
   \begin{center}
   \includegraphics[width=0.67\textwidth]{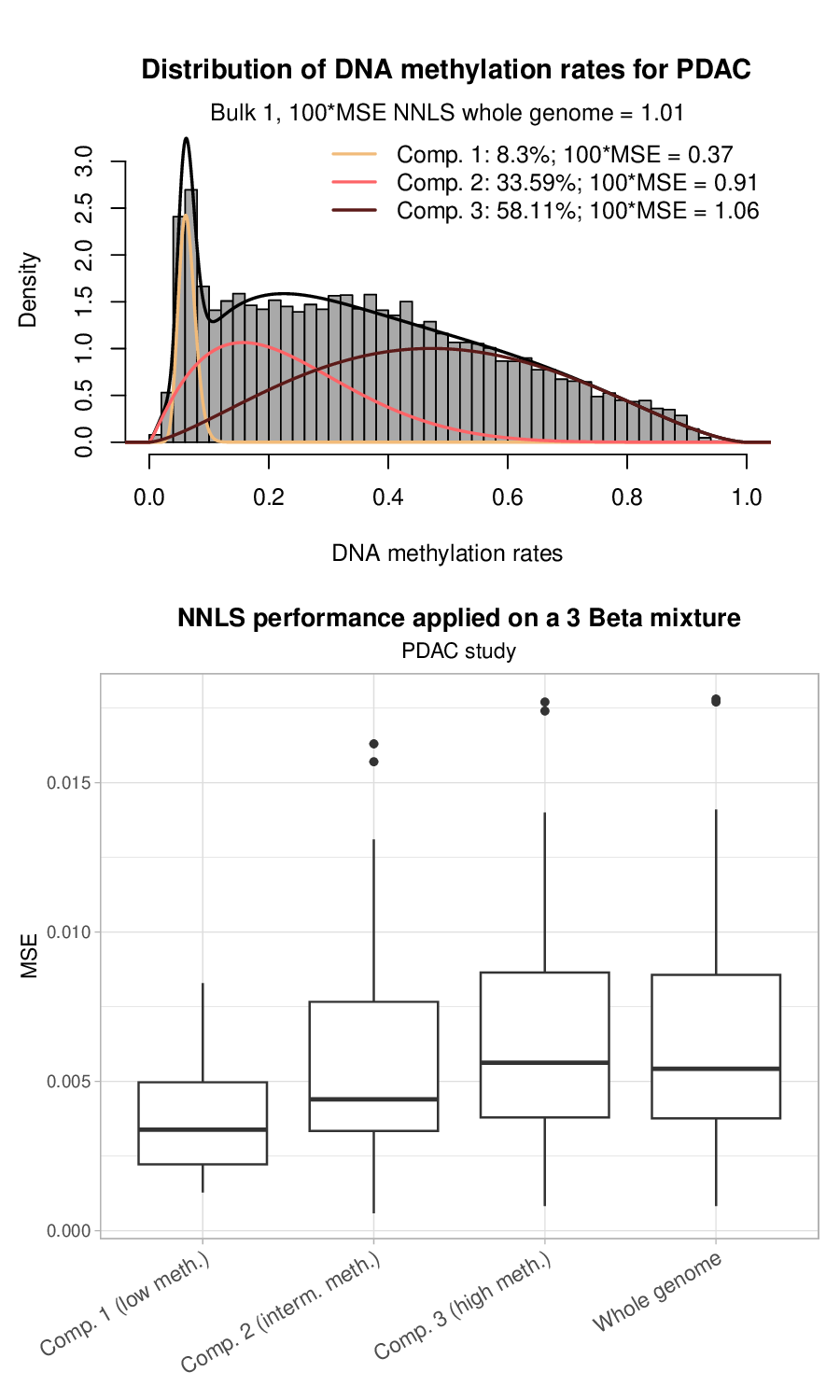}
   \end{center}
   \caption{PDAC study - Bulk 1 - Top plot: Histogram of the DNA methylation rates, with a three-component mixture of Beta distributions and their respective componentwise Mean Squared Error (MSE) of non-negative least squares (NNLS) estimation. Bottom plot: Mean Squared Error (MSE) of non-negative least squares (NNLS) estimation applied on each component and on the whole genome dataset.}
   \label{fig:3_beta_mixture_MSE}
\end{figure}

Mixtures of Beta distribution have already been used in other statistical genomics applications. Notably, three methylation states are frequently identified in methylation studies: hypomethylation, hypermethylation and intermediate values, potentially reflecting cellular heterogeneity or other biological mechanisms such as allele-specific or strand-specific methylation (hemimethylation). These methylation states are either defined with subjective threshold \cite{triche2016beta,chen2021filtering,men2017identification} or with mixture model \cite{majumdar2024_Rau_Beta_for_DNA,laurila2011betamixt_dim_reduc,gevaert2015methylmix,koestler2013recursively}. Consistently, a 3 component Beta mixture model is fitted to each bulk sample, with an illustration for Bulk 1 in Figure \ref{fig:3_beta_mixture_MSE}.

Applying the non-negative least squares (NNLS) algorithm separately to each of the three components identified in the mixture model of the DNA methylation rates yields markedly different estimates of the reference cell-type proportions. In the following, components are ranked according to their mean parameter: component 1 corresponds to the lowest mean (hypomethylated loci), component 2 to intermediate mean values, and component 3 to the highest mean (hypermethylated loci). Componentwise estimation of the cell type proportions is obtained by minimizing the weighted least squares criterion within the $p$-dimensional simplex ${\cal S}_{p}$, where the weights correspond to the posterior probabilities of component membership. For an illustrative purpose, supplementary Table \ref{tab:componentwise_cd_bulk_1} presents these heterogeneous estimates for Bulk 1, along with the corresponding MSE. 

The results shown in the bottom plot of Figure \ref{fig:3_beta_mixture_MSE} clearly indicate that restricting the analysis to genes belonging to component 1 leads to substantially more accurate estimates compared to deconvolution performed on the full set of genes or on subsets corresponding to components 2 and 3. This pattern is consistent across all bulk samples, suggesting that gene-level methylation rates associated with component 1, representing hypomethylated loci, may provide more informative signals or reduced noise for cell type proportion inference. This observation aligns with recent findings from \cite{guo2026guidelines_ref_panel}, who identified that prioritizing hypomethylated markers in the reference panel is one key factor for improving the accuracy of DNA methylation-based cell deconvolution.

Both the latent grouping structure observed in the deviance residuals and the existing literature provide strong motivation for employing a mixture of non-negative Beta regression models. Such models offer a flexible and insightful biologically realistic representation of bulk molecular data, with the number of components determined by standard model selection criteria in a data-driven and biologically agnostic manner. The resulting partition captures heterogeneous levels of model adequacy across the genome, reflecting the varying informativeness of genomic regions for cell-type deconvolution. We therefore propose a mixture of non-negative Beta regression (M-NNBR) model for cell deconvolution of bulk DNA methylation sample.

\subsection{Mixture of non-negative Beta regressions (M-NNBR)}
\label{s:Mixt-NNBR2}

Let $\mathcal{C} = (\mathcal{C}_{1},\dots,\mathcal{C}_{K})$ denote a $K$-dimensional latent class indicator, distributed as a multinomial $\mathcal{M}(1,\pi_{1},\dots,\pi_{K})$, where $K \ge 1$, $\pi_{k} = \mathbb{P}(\mathcal{C}_{k}=1) > 0$, and $\sum_{k=1}^{K} \pi_{k} = 1$. Conditionally on~$\mathcal{C}$, the bulk DNA methylation rates $y_{j}$, $j = 1,\dots,m$, follow a Beta distribution whose mean depends on the reference methylation profiles $x_{j}$ through a NNBR model:
\begin{eqnarray*} 
y_{j} | \{ \mathcal{C}_{jk} = 1\} \sim \mathcal{B}\bigl( \mu^{(k)}_{j} = \beta^{(k)}_{0} + x'_{j} \beta^{(k)}, \phi^{(k)} \bigr). 
\end{eqnarray*}
where $\beta_{0}^{(k)}$, $\beta^{(k)} = (\beta_{1}^{(k)},\ldots,\beta_{p}^{(k)})'$, with $\beta^{(k)}_{j} \geq 0$ for all $j=1, \ldots,p$, and $\phi^{(k)} > 0$ are respectively the component-specific intercept, regression coefficients, and precision parameters.

The unconditional density of $y_{j}$ is therefore given by the finite mixture:
\begin{eqnarray}  
f(y_{j} ; x, \pi, \beta_0, \beta, \phi) & = & \sum_{k = 1}^K \pi_k \varphi \bigl( y_{j} ; \mu_{j}^{(k)}, \phi^{(k)} \bigr), \label{eq:mixture_cd_density}
 \end{eqnarray}
where $\varphi(\cdot)$ denotes the Beta density specified in \eqref{eq:cellbetamod}. Writing $\beta_{0} = (\beta_{0}^{(1)},\dots,\beta_{0}^{(K)})'$, $\beta = (\beta^{(1)},\dots,\beta^{(K)})'$, and $\phi = (\phi^{(1)},\dots,\phi^{(K)})'$, the log-likelihood for model \eqref{eq:mixture_cd_density} takes the form
\begin{eqnarray} \label{eq:mixture_cd_LL} 
\mathcal{L}(\Theta) & = & \sum_{j=1}^m \text{log} \left( \sum_{k=1}^K \pi_k \varphi \bigl( y_j ; \mu_{j}^{(k)}, \phi^{(k)} \bigr) \right), 
\end{eqnarray}
where $\Theta = (\pi, \beta_{0}, \beta, \phi)'$ collects all model parameters.

Following \cite{grun2012extended}, an Expectation-Maximization (EM) algorithm is developed for maximum likelihood estimation of the proposed mixture of NNBR models, with \textit{a posteriori} enforcement of the sum-to-one constraint on the component-specific regression vectors. Because the bulk DNA methylation rates arise from convolution of the underlying cell-type-specific methylation signatures, their unconditional distribution naturally inherits a latent class structure. In line with this interpretation, the EM algorithm is initialized by fitting a mixture of Beta distributions to the bulk DNA methylation data. At the M-step, a weighted version of the GCD algorithm introduced above for whole-genome estimation of the NNBR model is used to maximize the component-specific weighted log-likelihoods. Full algorithmic details are provided in Appendix~\ref{app:mixture_em}.

We propose to determine the number $K$ of components in the M-NNBR model by minimization of the BIC (Bayesian Information Criterion), calculated with the componentwise non-negative estimates of the regression parameters, before applying the final sum-to-one normalization:
\begin{eqnarray} \label{eq:BIC} 
\text{BIC} \;=\; -2 {\cal L}(\hat{\Theta}) + \text{log} ( m ) \bigl( K(p+1) + p-1 \bigr),
\end{eqnarray}
where $\hat{\Theta}$ denotes the maximum-likelihood estimator of $\Theta$.

Applied to all bulk samples in the four benchmark studies used for method comparison in Section \ref{s:comp_stud}, BIC minimization selected M-NNBR models with between four and eight components. This range highlights the substantial heterogeneity of deconvolution models across the genome. The M-NNBR framework thus provides a flexible approach for partitioning the genome into gene sets that yield distinct and component-specific estimates of cell-type proportions. In Section \ref{s:gene_selection}, we introduce strategies for selecting a single gene set that offers the most accurate proportion estimates.

\section{Gene selection using M-NNBR}
\label{s:gene_selection}

In most statistical genomics applications, gene-level measurements (e.g., expression values) serve as variables whose relationships with a phenotype of interest are studied. Within this conventional framework, gene selection can be formulated as a variable selection problem and addressed using multiple-testing procedures in differential analysis or model-selection strategies in predictive modeling. Statistical criteria such as the false discovery rate or information criteria (e.g. BIC, AIC) then provide principled means of comparing alternative gene sets and identifying a minimal subset of relevant variables.

In the present cell deconvolution setting, genes are not treated as explanatory variables but rather as statistical units. The objective is to identify a subset of genes for which fitting the cell deconvolution model yields the most accurate estimates of cell-type proportions. Consequently, selecting an optimal gene set calls for \textit{ad hoc} criteria specifically designed to assess gene selection from an estimation-oriented perspective.

\begin{figure}[!ht]
  \begin{center}
  \includegraphics[width=\textwidth]{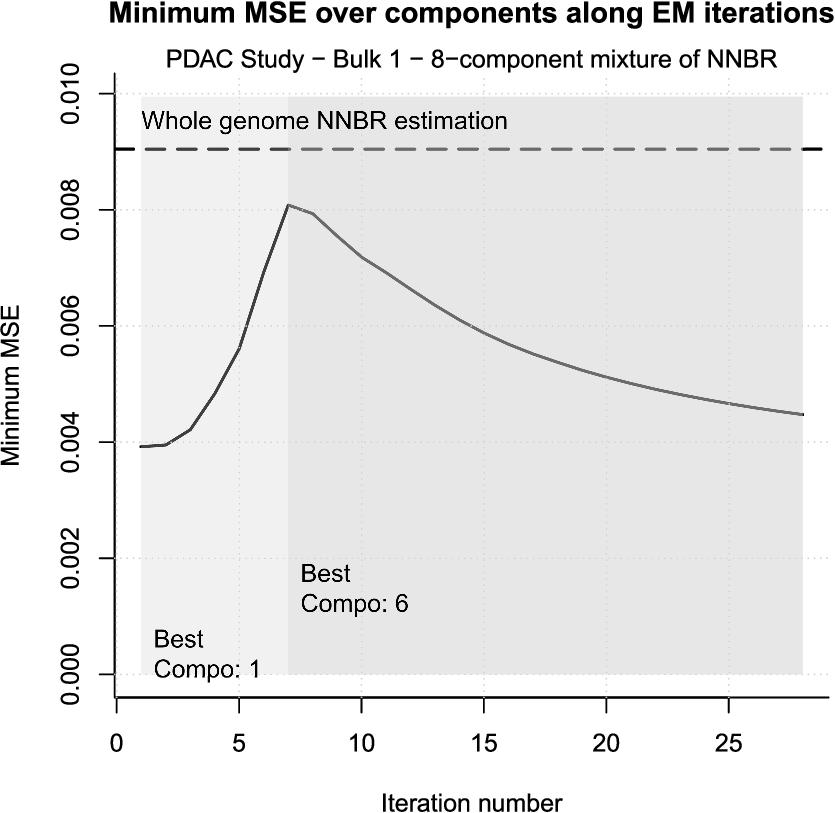}
  \end{center}
  \caption{PDAC study - Bulk 1 - M-NNBR model with eight components. Lowest Mean Squared estimation Error across components along iterations of the EM algorithm. The number of components is determined by minimization of the BIC.}
  \label{fig:minimum_mse_bulk1}
\end{figure}

The EM estimation of the M-NNBR model targets a globally adequate fit that captures heterogeneity in cell-type deconvolution models across the genome. However, it is not explicitly designed to optimize estimation accuracy within any single mixture component. This limitation is illustrated in Figure \ref{fig:minimum_mse_bulk1}, which displays the evolution, across EM iterations, of the minimum mean squared estimation error (MSE) over corresponding components when fitting an eight-component M-NNBR model to the DNA methylation data from Bulk 1 of the PDAC study. Notably, at each iteration, the best component-specific estimation accuracy exceeds that obtained from a whole-genome NNBR fit. During the early iterations, component 1 achieves the lowest MSE, whereas from iteration 7 onward and until convergence, component 6 provides the most accurate estimates. Overall, in this example, selecting the genome partition induced by the mixture model at the first EM iteration identifies the component that ultimately yields the best estimation accuracy.

Each iteration of the EM algorithm provides a partition of the genome into K components. The example discussed above (see Figure \ref{fig:minimum_mse_bulk1}) demonstrates that the final partition obtained at convergence of the algorithm is not necessarily the best one, in the perspective of identifying the component in which estimation accuracy is optimal. In the following, two implementations of the M-NNBR algorithm will be devised, the first one based on the partition obtained at the initial step of the EM algorithm and the second one at the last step, when the algorithms stops.

In line with most feature selection procedures in cell deconvolution, our component selection criterion targets numerical stability of estimation. Specifically, this criterion measures stability by the condition number of the asymptotic variance-covariance matrix of the component-specific vector of estimated cell-type proportions.

Closed-form expressions for these asymptotic variance-covariance matrices are available for the unconstrained beta regression model \cite{ferrari2004beta}. However, when non-negativity constraints are imposed on the parameters, as in the NNBR model, standard Fisher information-based asymptotic variances remain valid only when the true parameter lies in the interior of the parameter space \cite{self1987asymptotic}. In practice, it is frequently observed that some estimated proportions are equal or close to zero, placing the estimator near the boundary of the parameter space.

To mitigate this issue, we define, for each component, an active set of cell populations ${\cal A}_{k}$ as the smallest subset of cell types whose estimated proportions cumulatively exceed $0.95$. Focusing inference on this active set, we obtain the plug-in estimator $\hat{\Sigma}\!\left(\hat{\beta}_{k}\right)$ by substituting the maximum likelihood estimates into the expression of $\Sigma\!\left(\hat{\beta}_{k}\right)$ given by \cite{ferrari2004beta}, using the identity link function.

Supplementary Figure \ref{fig:S_bulk1_pdac} displays the asymptotic correlation matrices of the estimators of the cell-type proportions obtained from the NNBR model on Bulk 1, when fitted to the whole genome and when restricted to the genes assigned to component 1 at the initialization step. Gene assignment to components is performed using the standard maximum a posteriori (MAP) rule applied to the posterior probabilities estimated at the EM iteration selected in the first step of the procedure. The reported correlation matrices are directly derived from the corresponding asymptotic variance-covariance matrices, whose explicit expressions are provided in \cite{ferrari2004beta}.

In the whole-genome analysis, large correlations are observed between the estimated proportions of several pairs of cell populations as shown in left panel of the supplementary figure~\ref{fig:S_bulk1_pdac}, notably (\texttt{Macrophages}, \texttt{Neutrophils}), (\texttt{Cancer basal}, \texttt{Cancer classical}), (\texttt{CD4}, \texttt{CD8}), and (\texttt{Fibroblasts}, \texttt{Endothelial}). These strong correlations reflect the high similarity of DNA methylation rates within the corresponding pairs of reference cell populations and induce instability in the estimation of the proportion parameters, potentially impairing estimation accuracy. In contrast, right panel of Supplementary Figure \ref{fig:S_bulk1_pdac} shows that these correlations are reduced when the NNBR model is fitted exclusively to the genes belonging to component 1, highlighting the stabilizing effect of selecting this component on proportion estimation.

We propose to measure the componentwise stability of estimation in the NNBR model by the condition number of the variance-covariance matrix $\hat{\Sigma}(\hat{\beta}_{k}), \ k=1,\ldots,K,$:
\begin{eqnarray}
\kappa_{k} & = & \frac{\lambda_{\text{max}} ( \hat{\Sigma}(\hat{\beta}_{k}) )}{\lambda_{\text{min}} ( \hat{\Sigma}(\hat{\beta}_{k}) )}  ,   \label{eq:cw_cond_S}
\end{eqnarray}
where $\lambda_{\text{max}}(.)$ and $\lambda_{\text{min}}(.)$ provide respectively the largest and lowest eigenvalues of positive $p \times p$ matrices. An illustration of this componentwise stability one Bulk 1 is given in Appendix \ref{app:compo_stab_by_cond}.

In Section 5, the component selection criterion discussed above is compared to alternative standard gene selection methods used for cell deconvolution.

\section{Comparative studies}
\label{s:comp_stud}

Hereafter, we compare the proposed mixture of non-negative Beta regression deconvolution method (M-NNBR; Section \ref{s:Mixt-NNBR}) with the non-negative Beta regression (NNBR; Section \ref{s:Mixt-NNBR}) and four widely used deconvolution methods for DNA methylation data \cite{amblard2025robust,avila2020benchmarking,MethylDeconv_ferro2024}. These include: non-negative least squares (NNLS) \cite{nnls}, a constrained least squares method closely related to ordinary least squares (OLS); robust linear regression (RLR) \cite{MASS}, estimated via iteratively reweighted least squares, also referred to as Robust Partial Correlations (RPC) in the EpiDISH package \cite{EpiDISH_zheng2018}; MethylResolver \cite{arneson2020methylresolver} based on the least trimmed squares, similarly to FARDEEP \cite{fardeephao2019} used on bulk RNA-seq data ; and CIBERSORT \cite{MethylCIBERSORT_chakravarthy2018}, a widely used deconvolution method that has demonstrated consistent performance across both bulk RNA-seq and DNA methylation datasets. 

Among these methods, RLR, MethylResolver, and CIBERSORT belong to the family of robust regression approaches, designed to handle genomic heterogeneity and mitigate the influence of outlying observations through adaptive weighting or data partitioning strategies.

Each deconvolution method is evaluated on four publicly available benchmark datasets, using DNA methylation measurements aggregated at the gene level as described in Section \ref{s:background}:
\begin{itemize}
  \item \texttt{PDAC} (\cite{amblard2025robust}, introduce in Section \ref{s:background}): an \textit{in vitro} dataset comprising $p = 9$ cell types commonly observed in pancreatic ductal adenocarcinoma (PDAC) and $m = 18{,}735$ genes. This dataset was generated under strictly controlled experimental conditions, with known true proportions for each cell type in each of the $n = 30$ independent bulk samples. Two subgroups can be derived from each of the two PDAC cancerous subtype (classical and basal).
  \item \texttt{BlREAL2} \cite{reinius2012BlREAL2}: an \textit{in vivo} dataset comprising $p = 7$ distinct purified cell populations and $m = 18{,}865$ genes, with the reference matrix obtained by flow cytometry. It includes $n = 12$ bulk samples divided into two subgroups of six samples each: one consisting of PBMC samples with relatively balanced cell-type proportions, and the other consisting of whole-blood samples characterized by a dominant neutrophil fraction.
  \item \texttt{BlMIX2018} \cite{salas_BlMIX2018}: an \textit{in vitro} dataset comprising $p = 6$ distinct purified cell populations mixed in controlled proportions to generate $n = 12$ bulk samples, with $m = 24{,}638$ genes. As in \texttt{BlREAL2}, the samples can be divided into two subgroups of six samples each: one with relatively balanced cell-type proportions and one enriched in neutrophils, thereby more closely reflecting normal human peripheral blood.
  \item \texttt{BlMIX2016} \cite{koestler_BlMIX2016}: a related \textit{in vitro} benchmark dataset based on a similar experimental design. Because its structure and performance patterns were broadly comparable to those of \texttt{BlMIX2018}, the corresponding results are presented in Appendix \ref{app:BlMIX2016}.
\end{itemize}
Figure \ref{f:true_beta} in Appendix \ref{app:true_prop_heatmap} displays a heatmap of true proportions for each benchmark dataset.

Since the true cell-type proportions $\beta_j$, $j = 1, \ldots, p$, are known for all benchmark datasets considered in the comparison, deconvolution accuracy is assessed using the mean squared error (MSE), defined as
\[
\mathrm{MSE}(\hat{\beta}) = \frac{1}{p}\sum_{j=1}^{p}(\hat{\beta}_j - \beta_j)^2.
\]
In addition, because all methods are evaluated on the same bulk samples, we also consider the relative efficiency measure
\[
\mathrm{RE}(\hat{\beta}) = \frac{\mathrm{MSE}(\hat{\beta})}{\mathrm{MSE}(\hat{\beta}_{\footnotesize\text{NNLS}})},
\]
where $\mathrm{MSE}(\hat{\beta}_{\footnotesize\texttt{NNLS}})$ denotes the MSE of the NNLS deconvolution method, which is used here as the reference.

\bigskip

\noindent \textit{Gene selection using M-NNBR}

\bigskip

The proposed M-NNBR model induces a partition of the genome into a predefined number $K$ of components. This partition may be interpreted as soft, based on the posterior probabilities of component membership obtained from the EM algorithm, or as hard when each gene is assigned to a single component using a maximum a posteriori (MAP) rule. As described in Section \ref{s:Mixt-NNBR}, in the following implementations of M-NNBR, the number of components $K$ is selected by minimizing the Bayesian Information Criterion (BIC; equation \ref{eq:BIC}) evaluated at convergence of the EM algorithm. Given a partition, the task is to identify the component whose associated genes produce the most accurate cell deconvolution estimates.

As noted in Section \ref{s:gene_selection}, the partition obtained at convergence of the EM algorithm does not necessarily correspond to the component yielding the most accurate estimates. Alternative partitions include the initial partition derived from a mixture of Beta distributions fitted to the unconditional distribution of bulk DNA methylation rates, as well as all intermediate partitions generated throughout the iterations until convergence. Indeed, maximizing the likelihood of the mixture model may favor improved fit within larger components, even when the component achieving the highest estimation accuracy is smaller. This phenomenon is illustrated in Figure \ref{fig:minimum_mse_bulk1}, which displays the minimum MSE across mixture components over iterations for Bulk 1 of the \texttt{PDAC} study.

\begin{figure*}[!ht]
  \centerline{\includegraphics[width=\textwidth]{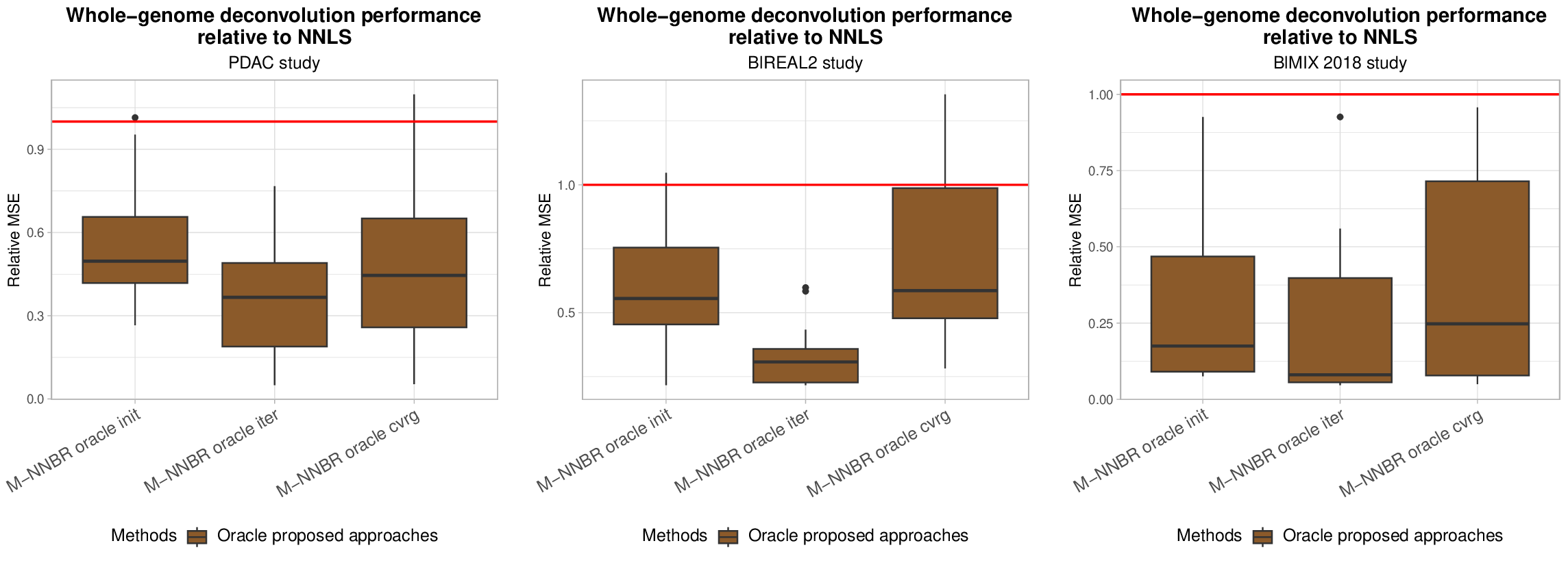}}
  \caption{Relative efficiencies (RE) with respect to NNLS across bulk samples in three benchmark studies (\texttt{PDAC}, \texttt{BlREAL2}, and \texttt{BlMIX2018}, from left to right). The component selected by the proposed mixture of non-negative Beta regression (M-NNBR) is chosen in an oracle manner for three types of partitions: at EM algorithm initialization (M-NNBR oracle init), at the iteration yielding the best estimation accuracy (M-NNBR oracle iter), and at convergence (M-NNBR oracle cvrg).} 
  \label{fig:mse_Mixt_NNBR_oracle_relative}
\end{figure*}

In all benchmark studies, Figure \ref{fig:mse_Mixt_NNBR_oracle_relative} presents boxplots of relative efficiencies across bulk samples for the best-performing component under three partitioning schemes: at initialization of the EM algorithm (M-NNBR oracle init), at the iteration for which the partition contains the component yielding the most accurate estimates (M-NNBR oracle iter), and at convergence (M-NNBR oracle cvrg). The best component is selected here in an oracle manner, solely to illustrate how the proposed mixture model approach can uncover informative partitions of the genome from a cell deconvolution perspective. 

Across all benchmark studies, the best component yields substantially more accurate estimates of true cell-type proportions in median than the whole-genome implementation of NNLS, regardless of the iteration at which the partition is defined. The best component identified at initialization also achieves accuracy comparable to or better than that at convergence, and remains close to the best performance observed across all iterations of the EM algorithm, with the exception of \texttt{BlREAL2}.

In the following, in order to avoid an additional iteration selection step, selection of the best component according to the procedure described in Section \ref{s:gene_selection} is performed on two partitions obtained using M-NNBR: (i) the initial partition resulting from fitting a mixture of Beta distributions to the unconditional distribution of bulk DNA methylation rates (M-NNBR-init), and (ii) the final partition at convergence (M-NNBR-cvrg). Notably, the initial partition exhibits a gradient in componentwise mean DNA methylation levels, whereas, across all bulk samples in the benchmark studies, the components in the final partition display relatively similar means and standard deviations. Moreover, goodness of fit is additionally enforced by requiring that the residual sum of squares of the selected component does not exceed that of the whole-genome NNBR model, preventing the selection of a component that is stable but poorly fitted.

\bigskip

\noindent \textit{Comparison of whole-genome cell deconvolution methods}

\bigskip

\begin{figure*}[!ht]
  \centerline{\includegraphics[width=\textwidth]{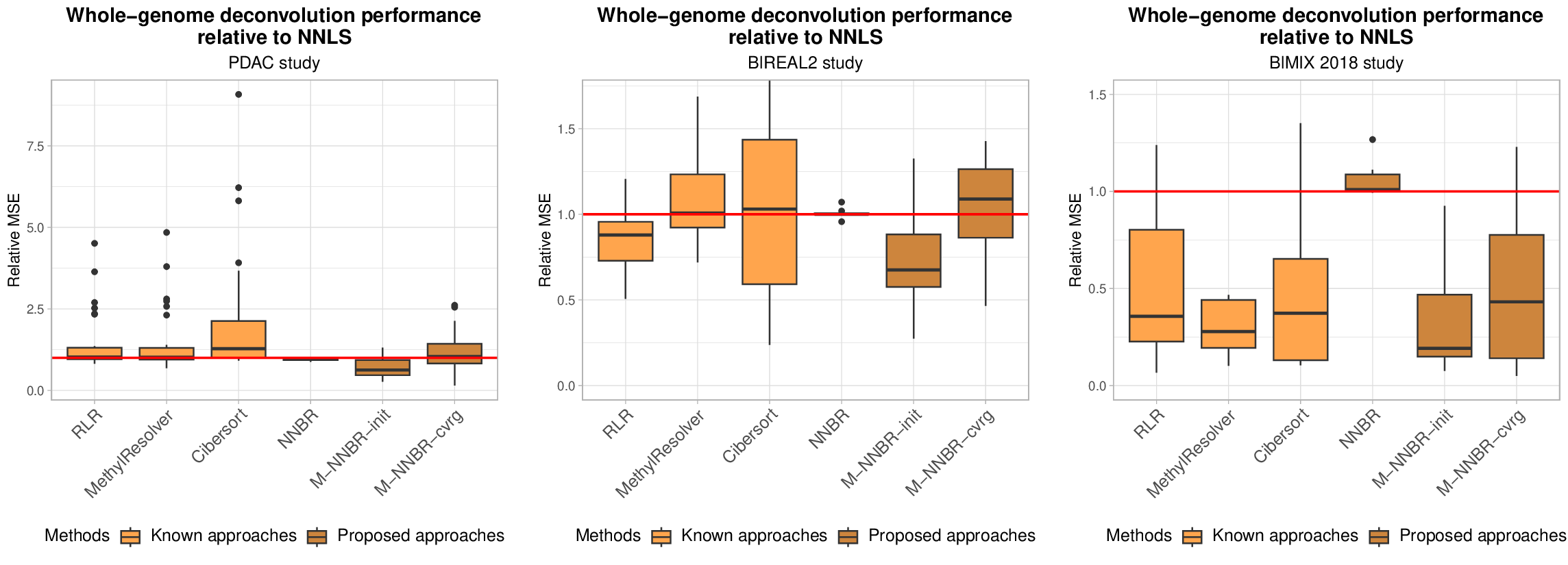}}
  \caption{Relative efficiencies (RE) of selected cell deconvolution methods with respect to NNLS across bulk samples in three benchmark studies (\texttt{PDAC}, \texttt{BlREAL2}, and \texttt{BlMIX2018}, from left to right). The proposed method is shown in two variants: M-NNBR-init and M-NNBR-cvrg, depending on whether the best component is selected from the partition at initialization of the EM algorithm or at convergence. Values above 1 (red horizontal line) indicate lower accuracy relative to NNLS.}
  \label{fig:mse_whole_genome_vs_Mixt_NNBR_relative}
\end{figure*}

Figure \ref{fig:mse_whole_genome_vs_Mixt_NNBR_relative} presents boxplots of relative efficiencies (RE) across bulk samples for all deconvolution methods in each benchmark study. All methods are evaluated on complete gene-aggregated methylation datasets without any prior feature selection, allowing comparison of their intrinsic whole-genome performance.

Across all benchmark studies, our approach with component selection at initialization (M-NNBR-init) achieves the lowest median RE. In the \texttt{PDAC} \textit{in vitro} dataset, where the best median RE is 0.627, none of the known deconvolution methods improves upon the performance of NNLS, including the M-NNBR-cvrg variant. 

In contrast, for the two blood-based benchmark datasets (\texttt{BlREAL2} and \texttt{BlMIX2018}), all methods achieve median estimation accuracy comparable to or better than NNLS. The best-performing approach, M-NNBR-init, attains a median RE of 0.675 and 0.192 for respectively \texttt{BlREAL2} and \texttt{BlMIX2018}, followed by RLR in \texttt{BlREAL2} (median RE: 0.879) and by MethylResolver in \texttt{BlMIX2018} (median RE: 0.278). These results also highlight the close similarity in performance between NNLS and NNBR, which yield comparable estimation accuracy across all datasets. This close similarity on whole-genome data suggests that replacing the Gaussian assumption with a Beta distribution is insufficient to improve deconvolution accuracy without a reduction of non-informative genomic regions.

The consistently lower performance of M-NNBR-cvrg relative to M-NNBR-init can be partly explained by less effective component selection at convergence. This may be attributed to the increased similarity of componentwise multivariate distributions of cell-type-specific DNA methylation levels in the final partition, which reduces the discriminative power for identifying the most informative component.

Across each of the benchmark studies considered for evaluating cell deconvolution methods, two distinct groups of bulk samples can be identified, each characterized by strong within-group similarity in cellular composition (see supplementary Figure \ref{f:true_beta}). Supplementary Figure \ref{fig:absolut_mse_by_subgroup} in Appendix reports the absolute mean squared error (MSE) of all cell deconvolution methods introduced above stratified by subgroup for each dataset.

In the \texttt{PDAC} dataset, one group of bulk samples, characterized by a higher proportion of basal-like cancer cells relative to classical cancer cells, exhibits lower median MSE across methods, suggesting that these samples are inherently easier to deconvolve. In contrast, for the subgroup enriched in classical cancer cells, M-NNBR-init improves deconvolution accuracy, thereby driving the overall performance gain observed for this dataset.

For the \texttt{BlMIX 2018} \textit{in vitro} and \texttt{BlREAL2} \textit{in vivo} datasets, the subgroup with more homogeneously distributed cell-type proportions is consistently more challenging to deconvolve than the subgroup dominated by a high proportion of neutrophils. In the \texttt{BlREAL2} dataset in particular, M-NNBR-init achieves the best median estimation accuracy within the more challenging subgroup, mirroring the pattern observed for \texttt{PDAC}. In contrast, RLR primarily drives overall performance gains by improving estimation accuracy in the neutrophil-dominated subgroup.

Furthermore, CIBERSORT exhibits relatively low variability in MSE across bulk samples within each subgroup of the two blood datasets, yet shows the largest performance discrepancy between subgroups in the \texttt{BlREAL2} dataset. The low variability of CIBERSORT within subgroup suggests that its internal feature selection responds consistently to the shared compositional structure of each subgroup. The large performance gap between the neutrophil-dominated and PBMC subgroups likely reflects the contrasting deconvolution difficulty, a pattern also observed in \texttt{BlMIX2018} but amplified in \texttt{BlREAL2}, possibly due to the additional biological variability inherent to \textit{in vivo} data.

\bigskip

\noindent \textit{Comparison after gene selection}

\bigskip

As introduced in section \ref{s:gene_selection}, feature selection in cell deconvolution consists in identifying a subset of informative genes or CpG sites, treated as statistical units. Such features may be selected based on prior biological knowledge, such as the known cell-type specificity of certain genes, or identified using statistical criteria designed to capture this specificity \cite{decamps_guidelines_2020,li2019TOASTFS,guo2026guidelinesFS}. By prioritizing highly cell-type-specific signals, this selection step reduces correlation among cell-type genomic profiles, yielding a near block-structured reference matrix that enhances the stability of deconvolution estimates.

Standard feature selection approaches in cell deconvolution rely exclusively on the reference matrix, which is shared across all bulk samples within a study. As a result, all bulk DNA methylation data associated to the same reference matrix are restricted to the same set of informative genes. In contrast, M-NNBR provides a principled framework for bulk-specific feature selection, whereby the most informative partition of the genome is identified independently for each bulk sample through the mixture structure.

In the following, we evaluate two commonly used feature selection procedures based on statistical measures of gene-specific variability across cell types. The first selects statistical units with the largest coefficient of variation of the vector $x=(x_1,\ldots,x_p)$, representing DNA methylation levels across the $p$ reference cell types, as implemented in the TOAST package \cite{li2019TOASTFS}. The second selects units with the largest variance of $x$ \cite{decamps_guidelines_2020}. Both procedures are applied as a preprocessing step to all considered deconvolution methods, namely NNLS, RLR, MethylResolver, CIBERSORT, and NNBR. Since feature selection is inherently embedded within the M-NNBR framework, no additional preliminary selection step is applied to this method.

A well-known limitation of such \textit{a priori} feature selection strategies is the need to specify the number of features to retain. This key hyperparameter is highly context-dependent, and no general consensus exists regarding its optimal choice \cite{guo2026guidelinesFS}, despite its substantial impact on deconvolution accuracy \cite{garmire2024challenges,MethylDeconv_ferro2024}. In M-NNBR, the number of selected genes arises naturally from the mixture model and corresponds to the size of the best-performing component within the inferred genomic partition.

In the comparative analyses below, the number of selected features is fixed across methods to ensure comparability. Specifically, it is set to the mean number of genes identified by maximum a posteriori in the best-performing mixture component, rounded to the nearest hundred: 1800 for \texttt{PDAC}, 1700 for \texttt{BlREAL2}, and 6000 for \texttt{BlMIX 2018}.

\begin{figure*}[!ht]
  \centerline{\includegraphics[width=\textwidth]{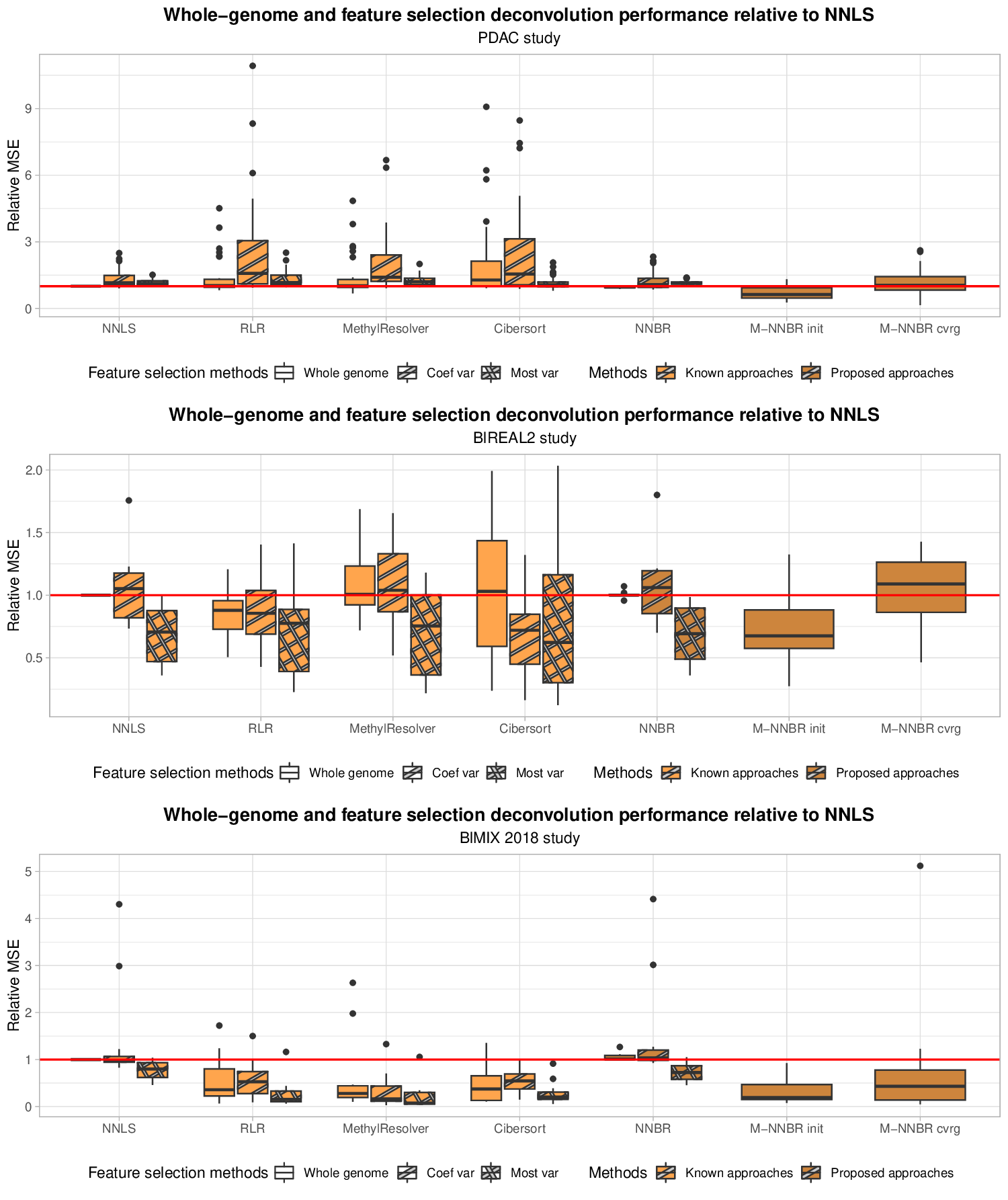}}
  \caption{Relative efficiencies (RE) of selected cell deconvolution methods with respect to NNLS across bulk samples from three benchmark studies (\texttt{PDAC}, \texttt{BlREAL2}, and \texttt{BlMIX2018}, from top to bottom). Two feature selection strategies, representative of commonly used statistical criteria in deconvolution pipelines, are evaluated for all known deconvolution approaches and NNBR: (i) selection based on the highest gene-wise coefficient of variation (Coef var) in the reference matrix, and (ii) selection based on the highest gene-wise variance (Most var).}
  \label{fig:mse_whole_genome_and_fs_vs_Mixt_NNBR_relative}
\end{figure*}

Figure \ref{fig:mse_whole_genome_and_fs_vs_Mixt_NNBR_relative} presents boxplots of relative efficiencies (RE) for all deconvolution methods across the benchmark studies, evaluated both on the full set of genes and after applying the feature selection strategies described above. Feature selection based on the coefficient of variation yields limited improvements, enhancing performance only for CIBERSORT in the \texttt{BlREAL2} dataset and for MethylResolver in \texttt{BlMIX 2018}. In both cases, however, selecting genes with the largest variance leads to comparable performance.  

Overall, variance-based feature selection substantially improves estimation accuracy for most methods in the \texttt{BlREAL2} and \texttt{BlMIX 2018} datasets. In particular, for \texttt{BlMIX 2018}, MethylResolver, RLR, and CIBERSORT applied to the variance-selected genes achieve the highest accuracy, with median RE values close to 0.15, comparable to the median RE of 0.19 obtained by M-NNBR-init. A similar pattern is observed for \texttt{BlREAL2}, where feature selection reduces performance disparities across methods, leading to more homogeneous results with median RE values around 0.7.  

In contrast, for the \texttt{PDAC} dataset, feature selection does not improve performance compared with whole-genome estimation. In this setting, M-NNBR remains the best-performing deconvolution approach.

As a perspective, biological prior knowledge could be incorporated into the M-NNBR component selection criterion, explicitly favoring components with strong cell-type specificity. For instance, differentially methylated regions or methylation markers could serve as the basis for a biologically informed selection criterion, directly linking the chosen genome partition to known cell-type-specific signatures. Beyond methylation data, hierarchical multi-omics integration \cite{picard2021integration} would presumably broaden the scope of M-NNBR, with component and iteration selection criteria informed by markers or regulatory knowledge from complementary omics studies.

\section{Conclusion}
\label{s:conclusion}

We developed a cell deconvolution model, M-NNBR, based on Beta regression, specifically designed to account for the bounded and asymmetric nature of DNA methylation rates. The heterogeneity of cell deconvolution model is handled by a latent group variable partitioning the genome into components with specific estimations of cell type proportions.

Latent group structures in genome-wide methylation data have previously been exploited through mixture modeling for various statistical objectives \cite{majumdar2024_Rau_Beta_for_DNA,laurila2011betamixt_dim_reduc,gevaert2015methylmix,koestler2013recursively}. In the context of cell deconvolution, however, our mixture of non-negative Beta regressions (M-NNBR) introduces a fundamentally different perspective on gene selection by linking it to bulk-specific model estimation. In contrast, standard feature-selection strategies rely exclusively on the reference matrix. Moreover, they require an arbitrarily chosen number of features to retain, a dataset-dependent hyperparameter without consensus guidance \cite{MethylDeconv_ferro2024}. M-NNBR requires no such choice, as the relevant gene set emerges directly from the mixture structure. This perspective also relates M-NNBR to the broader family of robust deconvolution approaches, such as RLR, MethylResolver, and CIBERSORT, which address genomic heterogeneity through implicit or explicit downweighting of genes identified to worsen model estimation.

Comparative analyses across several benchmark datasets demonstrate that M-NNBR consistently outperforms all competing methods when applied to whole-genome methylation data. This flexibility, however, comes with an additional methodological challenge: because the EM algorithm optimizes the global mixture likelihood rather than the estimation accuracy of individual components, identifying the most informative component for deconvolution becomes critical. To address this issue, we proposed a component selection criterion based on estimation stability, quantified through the condition number of the asymptotic variance-covariance matrix of the estimated cell-type proportions. The resulting procedure achieved stable and accurate performance across benchmark datasets spanning distinct biological contexts, suggesting that the proposed criterion captures a robust and transferable signal in genome-wide methylation data.

The comparison between M-NNBR and more classical pipelines combining \textit{a priori} feature selection with robust deconvolution methods reveals a more nuanced picture. Selecting genes with the most variable methylation levels substantially improves the performance of competing approaches, and both MethylResolver and CIBERSORT can slightly outperform M-NNBR on blood datasets when combined with this strategy. These results indicate that targeted feature selection remains highly effective when informative genes can be reliably identified beforehand. In contrast, both feature-selection strategies globally degrade performance on the PDAC \textit{in vitro} dataset, where M-NNBR remains the best-performing method. This observation suggests that the mixture framework may implicitly identify informative genomic regions in a more adaptive and context-robust manner than predefined feature-selection procedures.

Overall, the strong performance of M-NNBR across all benchmark datasets establishes mixture-based deconvolution as a promising methodological direction with substantial room for further improvement. In particular, while the EM initialization-based version of the method already captures a robust and generalizable deconvolution signal, component selection at later EM iterations remains an open problem. Moreover, the computational cost of this mixture of Beta regression is non-negligible and increases with the number of components and the size of the dataset. These analyses highlight two major perspectives for future work. First, component selection could likely be improved through more refined selection criteria or by incorporating prior biological knowledge. Second, adaptive selection of both mixture components and EM iteration may further enhance deconvolution accuracy. Altogether, our results identify the detection of informative gene subsets and the development of principled \textit{a posteriori} component-selection strategies as the key remaining challenges for mixture-based cell deconvolution.

\section*{Author Contributions}

H.B. and D.C. were responsible for conceptualization and methodology. H.B. was responsible for formal analysis, visualization, and writing the original draft. Y.B., M.R., and D.C. were responsible for supervision. All authors contributed to reviewing and editing of the manuscript.

\section*{Conflicts of Interest}

The authors declare no conflicts of interest.

\bibliographystyle{plain}
\bibliography{biblio}

\appendix

\section{Proportion estimation by NNLS and NNBR for Bulk 1 of the PDAC study}

For an illustrative purpose using the DNA methylation data for Bulk 1 of the \texttt{PDAC} study, Table \ref{tab:true_beta_bulk_1} displays respectively the cellular heterogeneity ground truth, estimation from NNLS and estimation from NNBR. Furthermore, Table \ref{tab:componentwise_cd_bulk_1} shows componentwise NNLS estimates apply on Bulk 1 from the mixture of 3 Beta distributions for the distribution of DNA methylation rates presented in Figure \ref{fig:3_beta_mixture_MSE}.

\begin{table}[!ht]
  \centering
  \caption{\texttt{PDAC} study - Bulk 1 - True proportions ($\%$) of nine reference cell populations (1st column) compared to NNLS and NNBR estimations (respectively 2nd and 3rd column).}
  \label{tab:true_beta_bulk_1}
  \begin{tabular}{lrrr}   
  \toprule
  Cell population    & Ground truth ($\%$) & NNLS ($\%$) & NNBR ($\%$)   \\  
  \midrule
  B cells            &    0.50                 &  0.00                 &  0.00   \\
  Cancer basal       &   20.63                 & 24.36                 & 25.01   \\
  Cancer classical   &    5.01                 & 12.07                 & 11.60   \\
  CD4 T cells        &    1.30                 &  0.00                 &  0.00   \\
  CD8 T cells        &    3.91                 &  0.00                 &  0.00   \\
  Endothelial        &   29.76                 & 47.73                 & 46.39   \\
  Fibroblasts        &   38.39                 & 15.84                 & 17.00   \\
  Macrophages        &    0.20                 &  0.00                 &  0.00   \\
  Neutrophils        &    0.30                 &  0.00                 &  0.00   \\   \hline
  MSE ($\times 10^{-2}$)  &    -               & 1.01                  & 0.90   \\   
  \bottomrule
  \end{tabular}
\end{table}

\begin{table*}[!t]
   \caption{\texttt{PDAC} study - Bulk 1 - Componentwise NNLS estimations of the proportions ($\%$) of nine reference cell populations. The three components are deduced from the mixture of Beta distributions for the distribution of DNA methylation rates (see Figure \ref{fig:3_beta_mixture_MSE})}
   \label{tab:componentwise_cd_bulk_1} 
   \begin{center}
   \resizebox{\textwidth}{!}{
   \begin{tabular}{lrrrr}
   \toprule
   Cell population    & Ground truth ($\%$) & Component 1 ($\%$) & Component 2 ($\%$) & Component 3 ($\%$)   \\  
   \midrule
   B cells            &    0.50                 &  0.00       &  0.40       &  0.00   \\
   Cancer basal       &   20.63                 & 13.93       & 26.39       & 24.02   \\
   Cancer classical   &    5.01                 &  4.92       & 12.17       & 11.92   \\
   CD4 T cells        &    1.30                 &  0.00       &  0.00       &  0.00   \\
   CD8 T cells        &    3.91                 &  0.00       &  0.00       &  0.00   \\
   Endothelial        &   29.76                 & 38.07       & 44.75       & 48.51   \\
   Fibroblasts        &   38.39                 & 30.82       & 16.29       & 15.55   \\
   Macrophages        &    0.20                 & 12.27       &  0.00       &  0.00   \\
   Neutrophils        &    0.00                 &  0.00       &  0.00       &  0.00   \\   \hline
   MSE ($\times 10^{-2}$) & -                   & 0.37        & 0.91        & 1.06    \\   
   \bottomrule
   \end{tabular}
   }
   \end{center}
\end{table*}

\section{Heteroscedasticity and deviance residual diagnostics for Bulk 1 of the PDAC study}

\begin{figure}[!ht]
   \begin{center}
   \includegraphics[width=\textwidth]{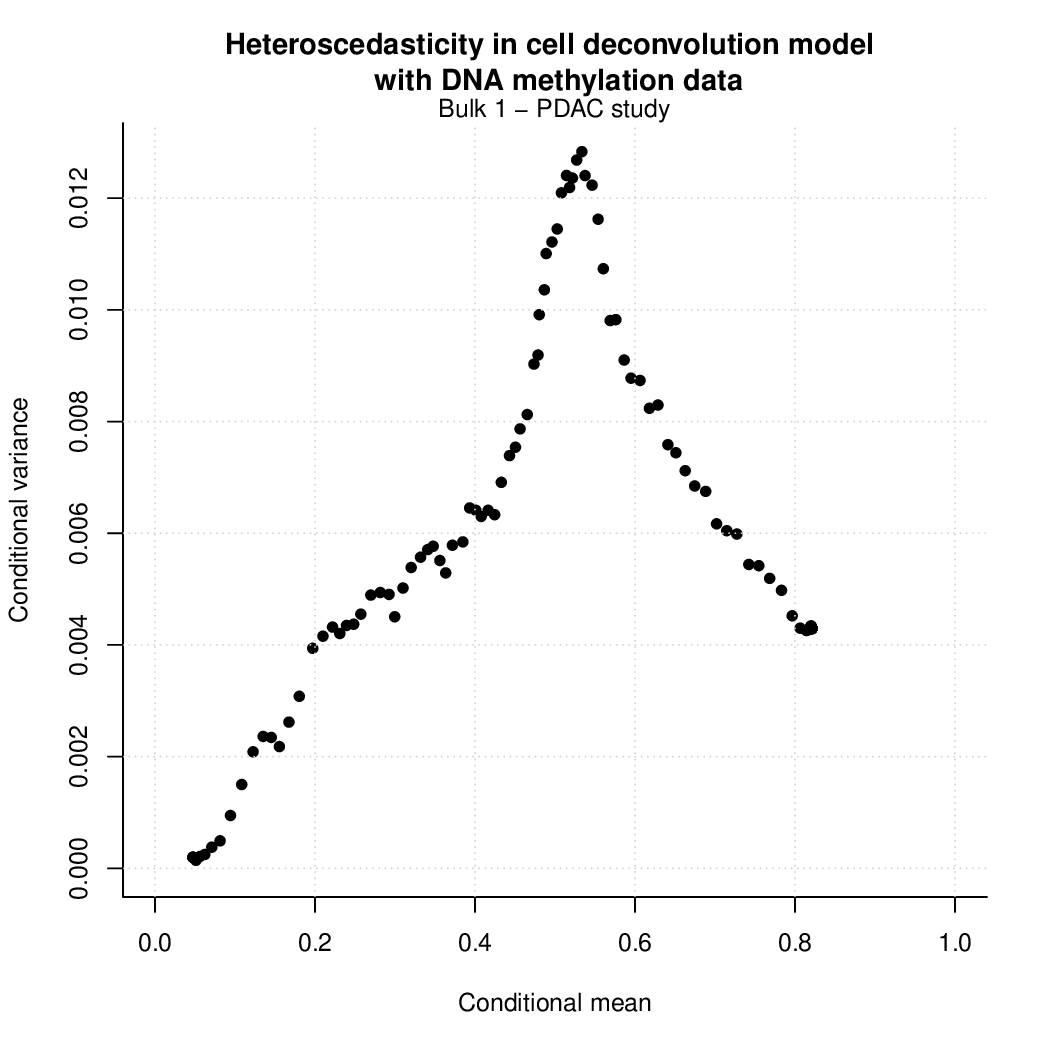}
   \end{center}
   \caption{\label{fig:overdispersion_bulk_1} \texttt{PDAC} study - Bulk 1 - Relationship between conditional variances and means of the bulk DNA methylation rates given the DNA methylation rates in the reference cell populations.}
\end{figure}

For an illustrative purpose using the DNA methylation data for Bulk 1 of the \texttt{PDAC} study, Figure \ref{fig:overdispersion_bulk_1} displays the empirical relationship between conditionnal variances and means, respectively approximate by local estimates $\tilde{m}(x)$ and $\tilde{v}(x)$. These estimates are computed for 100 evenly spaced values of $u(x)$, using local neighborhoods defined as the 5\% nearest genes in terms of $u(x_j)$.

\begin{figure}[!ht]
   \begin{center}
   \includegraphics[width=\textwidth]{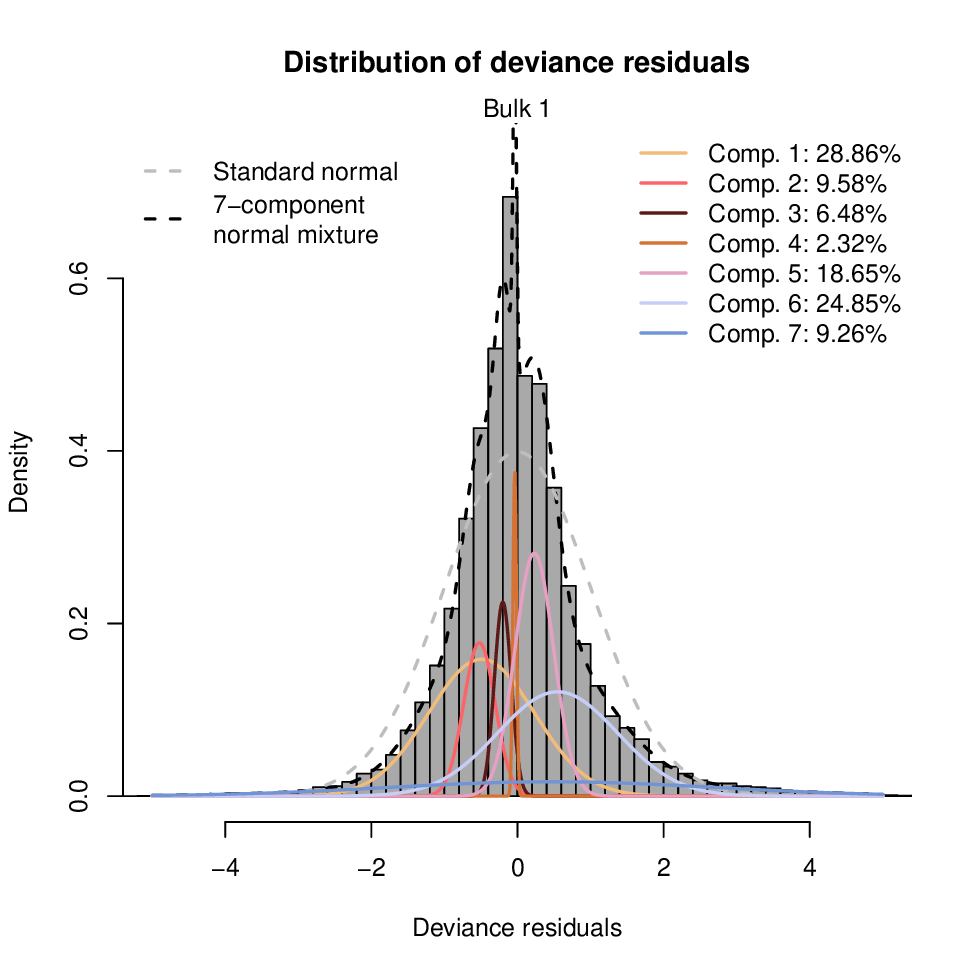}
   \end{center}
   \caption{\texttt{PDAC} study - Bulk 1 - Histogram of deviance residuals produced by the maximum-likelihood fit of the NNBR model \eqref{eq:cellbetamod}. The dashed grey curve corresponds to the standard normal density, and the dashed black curve to a seven-component normal mixture density. Componentwise mixture densities are shown as coloured solid lines. The number of mixture components is selected by minimization of the Bayesian Information Criterion (BIC).}
   \label{fig:deviance_residuals}
\end{figure}

In Figure \ref{fig:deviance_residuals}, the seven-component normal mixture is obtained using the \texttt{R} function \texttt{densityMclust} from the \texttt{mclust} package \cite{scrucca2023model}.

\section{Greedy coordinate descent estimation of the non-negative Beta regression model\label{app:gcd}}

We propose to estimate the NNBR model \eqref{eq:cellbetamod} using a greedy variant of the Cyclic Coordinate Descent (CCD) algorithm to maximize the log-likelihood function \eqref{eq:unweighted_LL}, subject to the constraint that all coordinates of $\beta$ are non-negative.

Initial estimates $\hat{\beta}^{(0)}_{0}$ and $\hat{\beta}^{(0)}$ for $\beta_{0}$ and $\beta$, respectively, can be obtained using any non-negative least-squares (NNLS) algorithm, such as the Lawson–Hanson method \cite{lawson1995solving}, implemented in the \texttt{R} package \texttt{nnls} \cite{nnls}.

Let $\hat{\mu}^{(0)} = \hat{\beta}^{(0)}_{0} \mathbf{1}_{m} + X \hat{\beta}^{(0)}$ denote the $m$ fitted values of the linear predictor obtained from the current estimates of $\beta_{0}$ and $\beta$, where $\mathbf{1}_{m}$ is an $m$-dimensional vector of ones and $X$ is the $m \times p$ signature matrix whose $j$th row is $x_{j}$.  
An initial estimate $\hat{\phi}^{(0)}$ of $\phi$ is then obtained by equating the first derivative of the log-likelihood function \eqref{eq:unweighted_LL} with respect to $\phi$, after substituting $\hat{\beta}^{(0)}_{0}$ and $\hat{\beta}^{(0)}$:
\begin{eqnarray*}
   \frac{\partial {\cal L}}{\partial \phi} ( \varphi, \hat{\mu}^{(0)} ; y , x ) & = & m \psi ( \phi ) - \sum_{j=1}^{m} \hat{\mu}^{(0)}_j \psi ( \hat{\mu}^{(0)}_j \phi ) -  \\
   & & \sum_{j=1}^{m} ( 1 - \hat{\mu}^{(0)}_j ) \psi ( ( 1 - \hat{\mu}^{(0)}_j ) \phi ) +  \\
   & & \sum_{j=1}^{m} \hat{\mu}^{(0)}_j \text{log} ( y_{j} ) + \sum_{j=1}^{m} ( 1 - \hat{\mu}^{(0)}_j ) \text{log} ( 1 - y_{j} )  ,  
\end{eqnarray*}
where $\psi(\cdot)$ denotes the digamma function, i.e., the first derivative of $\log \Gamma(\cdot)$.  
The root of this equation can be obtained using the \texttt{R} function \texttt{uniroot} implementing Brent's method \cite{brent1973some}.

Alternatively, as also mentioned by \cite{ferrari2004beta}, the following approximation of the digamma function $\psi(z)$ for large values of $z$ \cite{ronning1989maximum} can be used to speed up calculations:
\begin{eqnarray}
\psi(z) & \approx & \text{log} (z) - \frac{1}{2z} .    \label{eq:digamma_approx}
\end{eqnarray}

Using this approximation, the derivative of the log-likelihood with respect to $\phi$ can be approximated by:
\begin{eqnarray*}
   \frac{\partial {\cal L}}{\partial \phi} ( \varphi, \hat{\mu}^{(0)} ; y , x ) & \approx & \frac{m}{2\phi} +
   \sum_{j=1}^{m} \text{log} ( 1-y_{j} ) - \text{log} ( 1-\hat{\mu}^{(0)}_{j} ) +  \\
   & & \sum_{j=1}^{m} \hat{\mu}^{(0)}_{j} \Bigl(  \text{logit} ( y_{j} ) - \text{logit} ( \hat{\mu}^{(0)}_{j} ) \Bigr) .  
\end{eqnarray*}

Equating the above approximated derivative to zero leads to the following closed-form initial estimation of $\phi$:

\begin{align}
    D^{(0)} & = & \sum_{j=1}^{m} \Bigl( \log(1 - \hat{\mu}^{(0)}_j) 
              - \log(1 - y_{j}) \Bigr) \notag \\
            & + & \sum_{j=1}^{m} \hat{\mu}^{(0)}_j 
              \Bigl( \operatorname{logit}(\hat{\mu}^{(0)}_j) 
              - \operatorname{logit}(y_j) \Bigr), 
              \label{eq:def_D} \\[6pt]
    \hat{\phi}^{(0)} & \approx & \frac{m}{2 \, D^{(0)}}. 
    \label{eq:approx_phi}
\end{align}

The score function with respect to the regression parameters $(\beta_{0}, \beta')$ is now obtained by differentiating the log-likelihood with respect to these parameters:
\begin{eqnarray*}
{\cal S} ( \beta_{0} , \beta ; \phi )
= \phi
\begin{pmatrix}
\mathbf{1}_{m}' \\
X'
\end{pmatrix}
e(\beta_{0} , \beta ; \phi),
\end{eqnarray*}
where $e(\beta_{0}, \beta; \phi)$ is the $m$-dimensional vector with components
\begin{eqnarray*}
  e_{j} (\beta_{0} , \beta ; \phi) & = & - \big[ \psi ( \mu_j \phi ) - \psi ( ( 1 - \mu_j ) \phi ) \big] +  \\
  & & \big[ \log ( y_{j} ) - \log ( 1 - y_{j} ) \big] ,
\label{eq:nnbr_residuals}
\end{eqnarray*}
where $\mu_{j}=\beta_{0}+x'_{j}\beta$.

At iteration $t+1$, the estimate of $\beta_{0}$ is updated by setting the first coordinate of the score function to zero, after plugging-in the current estimates $\hat{\beta}^{(t)}$ and $\hat{\phi}^{(t)}$.

In the standard CCD maximization scheme, each non-negative coordinate $\beta_{j}$, $j = 1, \ldots, p$ of $\beta$ is updated sequentially by equating to zero the corresponding coordinate ${\cal S}_{j}(\cdot)$ of the score function, using the current estimates $\hat{\beta}^{(t)}_{0}$, $\hat{\beta}^{(t)}_{j'}$ (for $j' \neq j$), and $\hat{\phi}^{(t)}$.  
To ensure non-negativity, if the marginal maximization yields a negative value, the corresponding parameter is set to zero. Furthermore, if the current fitted value $\hat{\mu}^{(t)} = \hat{\beta}^{(t)}_{0} + x_{j}' \hat{\beta}^{(t)}$ lies outside the open interval $(0, 1)$, the corresponding gene $j$ is excluded from subsequent iterations.

Instead of cycling through all coordinates, convergence can be accelerated by updating, at each step, only the coordinate that yields the largest improvement in the log-likelihood.  
The algorithm terminates when the relative change in the log-likelihood between two successive iterations falls below a predefined tolerance threshold.

To further reduce computational time, for large values of $\phi$ we use the following approximation of $e_{j}(\beta_{0}, \beta; \phi)$ \cite{ferrari2004beta} deduced from approximation \eqref{eq:digamma_approx} of the \texttt{digamma} function:
\begin{eqnarray*}
\tilde{e}_{j} (\beta_{0}, \beta)
= - \big[ \log ( \mu_j ) - \log ( 1 - \mu_j ) \big]
+ \big[ \log ( y_{j} ) - \log ( 1 - y_{j} ) \big].
\label{eq:nnbr_residuals_approx}
\end{eqnarray*}

With this approximation, the updates of $\beta_{0}$ and $\beta$ no longer depend on the current estimate of $\phi$, thereby improving the stability and speed of convergence. In our implementation, $e_{j}(\beta_{0}, \beta; \phi)$ is approximated by $\tilde{e}_{j}(\beta_{0}, \beta)$ when $\phi \geq 200$. 

Finally, the updated estimate $\hat{\phi}^{(t+1)}$ is obtained using the same procedure as for initializing the algorithm: either by equating the exact first-order derivative of the log-likelihood with respect to $\phi$ to zero, or by using the closed-form approximation \eqref{eq:approx_phi} derived from \eqref{eq:digamma_approx} when $\hat{\phi}^{(t)}$ becomes large (greater than 200 in our implementation), evaluated at the current values of $\hat{\mu}^{(t)}$ rather than $\hat{\mu}^{(0)}$.

The GCD algorithm described above is applied to all bulk samples from the four studies introduced in Section \ref{s:comp_stud}. For comparison, a Fisher scoring algorithm is also implemented for the unconstrained estimation of the regression parameters. The non-negativity and sum-to-one constraints are then imposed \textit{a posteriori} by setting negative estimates to zero and renormalizing the resulting vector of non-negative estimates to sum to one. Because our beta regression model for cell deconvolution requires the identity link function, the exclusion rule described previously is also enforced at each iteration for the Fisher scoring procedure, removing genes whose fitted mean $\mu(x)$ falls outside the interval $(0,1)$. For both algorithms, the initial estimates of $\beta_{0}$ and $\beta$ are provided using the Lawson-Hanson implementation of the NNLS method implemented in the \texttt{R} package \texttt{nnls}. Convergence is declared when the relative change in successive log-likelihood values falls below $10^{-3}$.

\begin{figure*}[!t]
  \begin{center}
  \includegraphics[width=\textwidth]{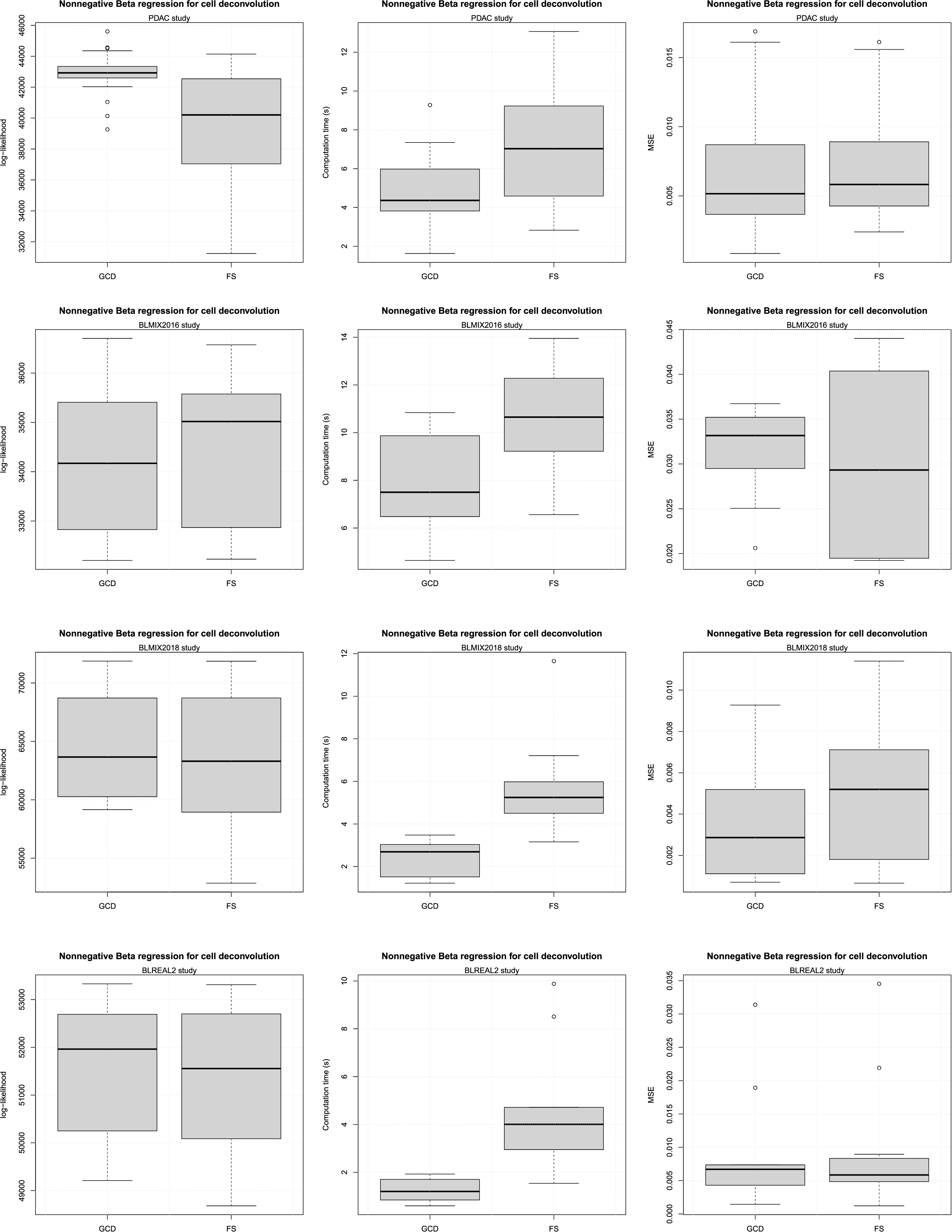}
  \end{center}
  \caption{\label{fig:app_gcd_versus_fs} Maximum achieved log-likelihood (Left column), computation time in seconds (middle column) and MSE (right column) for the estimation of the cell-type proportions in all bulks of the \texttt{PDAC} (1st row), \texttt{BLMIX2016} (2nd row), \texttt{BLMIX2018} (3rd row) and \texttt{BLREAL2} (4th row) studies introduced in Section \ref{s:comp_stud} using a non-negative beta regression model. GCD: Greedy coordinate descent. FS: Fisher scoring}
\end{figure*}

Figure \ref{fig:app_gcd_versus_fs} presents boxplots comparing the two estimation algorithms in terms of their maximum achieved log-likelihood, computation time, and mean squared error (MSE). The MSE is computed using the known true cell-type proportions for each bulk sample. Overall, the maximum log-likelihoods obtained with the two approaches are similar across studies, with the exception of the PDAC dataset, for which the GCD algorithm attains distinctly higher log-likelihood values across the 30 bulks in this study. The distributions of MSEs show no substantial differences between the two methods. In contrast, computation times differ markedly: the GCD algorithm is consistently and substantially faster than the Fisher scoring algorithm. Finally, for two bulk samples in the \texttt{BLMIX2016} study, the Fisher scoring procedure failed to produce maximum-likelihood estimates due to singularities in the matrix of second-order derivatives of the log-likelihood.
      
\section{Expectation-Maximisation algorithm for the mixture of non-negative Beta regression models\label{app:mixture_em}}

A standard approach for maximizing the log-likelihood in equation \eqref{eq:mixture_cd_LL} with respect to all model parameters is the Expectation-Maximization (EM) algorithm. The EM procedure alternates between an E-step (expectation) and an M-step (maximization) until convergence. Initialization can consist of first fitting a mixture model of Beta distributions for the unconditional distribution of bulk DNA methylation rates and then estimating componentwise cell deconvolution models.

Let $\hat{\theta}^{(l)}$ denote the estimate of $\theta$ after $l$ iterations. At iteration $l+1$, the E-step computes the $\mathcal{Q}$-function, defined as the conditional expectation of the complete-data log-likelihood given the observed data and the current parameter estimates:
$ \mathcal{Q}\!\left(\theta \,;\, \hat{\theta}^{(l)}\right) \;=\; \sum_{j=1}^m \sum_{k=1}^K \omega_{jk}^{(l)} \, \varphi\!\left( y_j \,;\, \mu^{(k)}(x_j), \phi^{(k)} \right),$
where
\begin{eqnarray*}
  \omega_{jk}^{(l)} &=& \frac{\hat{\pi}_k^{(l)} \, \varphi\!\left( y_j \,;\, \hat{\mu}^{(k),(l)}(x_j), \hat{\phi}^{(k),(l)} \right)}
  {\sum_{k'=1}^K \hat{\pi}_{k'}^{(l)} \,\varphi\!\left( y_j \,;\, \hat{\mu}^{(k'),(l)}(x_j), \hat{\phi}^{(k'),(l)} \right)},
\end{eqnarray*}
denotes the posterior probability at iteration $l$ that the methylation value $y_j$ originates from component $k$.

The M-step updates $\hat{\theta}$ by maximizing $\mathcal{Q}\!\left(\theta \,;\, \hat{\theta}^{(l)}\right)$ with respect to all model parameters. Because all component-specific parameters are distinct, this reduces to fitting $K$ separate weighted cell deconvolution models, where the weights are given by the posterior probabilities $\omega_{k}^{(l)} = (\omega_{1k}^{(l)}, \ldots, \omega_{mk}^{(l)})$. These model fits are obtained by maximizing the weighted log-likelihood formulation of the non-negative beta regression model (see equation~\ref{eq:unweighted_LL}), $\mathcal{L} ( \phi, \beta_{0}, \beta ; \omega )=\sum_{j=1}^{m} \omega_{j} \ell_{j} ( \phi, \beta_{0}, \beta ; \omega )$, using the GCD algorithm presented in Appendix \ref{app:gcd}.

 The mixture weights are then updated explicitly by averaging the posterior probabilities:
\begin{eqnarray*}
  \hat{\pi}_k^{(l+1)} &=& \frac{1}{m} \sum_{j=1}^m \omega_{j}^{(k),(l)} .
\end{eqnarray*}

The sum-to-one constraint on the component-specific vectors of estimated regression parameters is imposed \textit{a posteriori} by dividing each vector of non-negative estimates by its sum.

\section{Comparison of estimation stability in the PDAC study\label{app:compo_stab_by_cond}}

For an illustrative purpose using the DNA methylation data for Bulk 1 in the \texttt{PDAC} study, Table \ref{tab:cond_bulk1} displays the metric \eqref{eq:cw_cond_S} for estimation stability in the NNBR model fitted on each of the eight components at the first iteration of the M-NNBR model.            

\begin{table*}[!t]
\begin{center}
\caption{\label{tab:cond_bulk1} \texttt{PDAC} study - Bulk 1 - Component-specific stability metric $\kappa_{k}$ (see equation \ref{eq:cw_cond_S}) for M-NNBR model with 8 components at the first EM iteration. No gene are assigned to component 8, when using The MAP rule applied on the posterior probabilities estimated at the first EM iteration.}
\begin{tabular}{lrrrrrrrr} \toprule
 & \multicolumn{8}{l}{Component}    \\   \cline{2-9}
 & 1 & 2 & 3 & 4 & 5 & 6 & 7 & 8  \\   \midrule
 $\kappa_{k}$ & 210.43 & 263.87 & 782.54 & 1150.34 & 818.11 & 541.43 & 245.94 & - \\   \midrule
 Prior probability & 0.17 & 0.05 & 0.20 & 0.29 & 0.13 & 0.09 & 0.08 & 0.01 \\   
 RE ($\times 100$) & 0.43 & 0.67 & 0.83 & 1.22 & 1.26 & 1.19 & 1.20 & 1.42  \\  \bottomrule
\end{tabular}
\end{center}
\end{table*}  

\begin{figure*}[!t]
  \begin{center}
  \includegraphics[width=\textwidth]{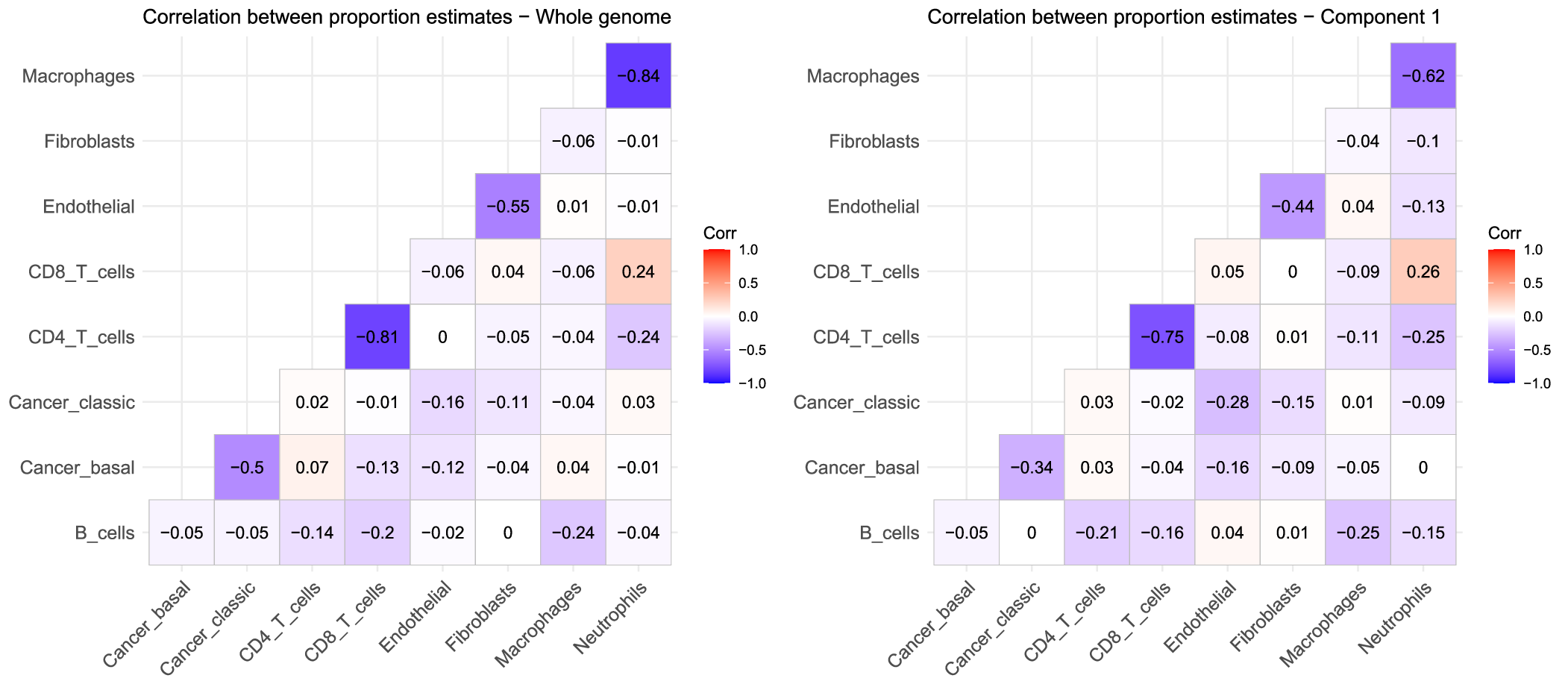}
  \end{center}
  \caption{\label{fig:S_bulk1_pdac} \texttt{PDAC} study - Bulk 1 - Asymptotic correlation matrices for estimators of the proportion parameters in the NNBR model, fitted to whole genome DNA methylation rates data (left plot) and to the genes associated with component 1 by MAP rule of the M-NNBR model.}
\end{figure*}

In the specific case of Bulk 1 in the \texttt{PDAC} study, component~1, which achieves the highest estimation accuracy, also exhibits the smallest values of the metric $\kappa_{k}$. Consequently, Figure \ref{fig:S_bulk1_pdac} displays the asymptotic correlations matrices for estimators of the M-NNBR respectively fitted to the whole genome (left plot) and to the genes in component 1 of Bulk 1 (right plot).

\section{True Proportion Heatmaps and Subgroup Estimation Across Benchmark Datasets\label{app:true_prop_heatmap}}

\begin{figure*}[!t]
  \begin{center}
  \includegraphics[width=\textwidth]{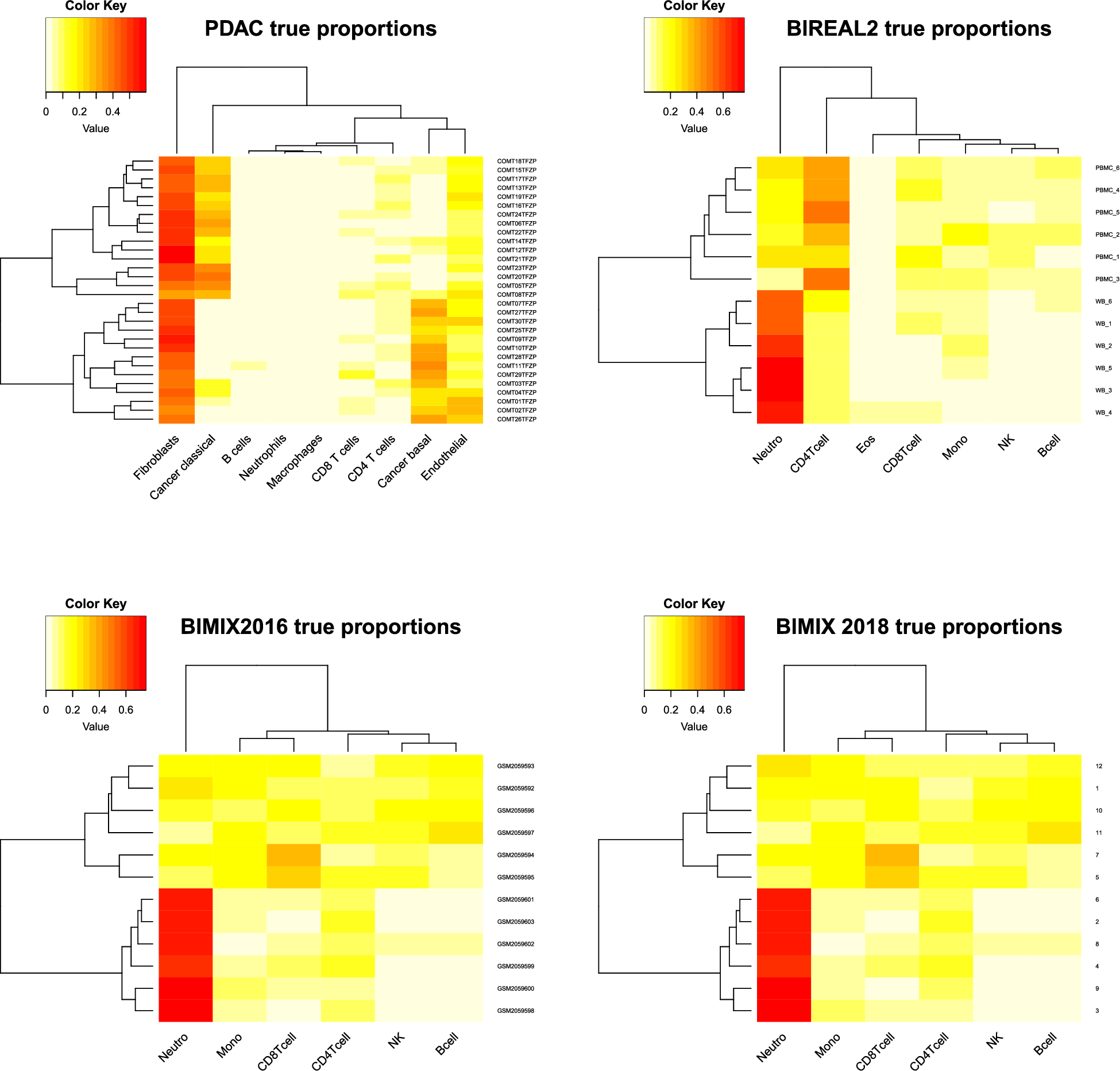}
  \end{center}
  \caption{\label{f:true_beta}Heatmap of the true proportions of each cell types (columns) within the bulks (rows) for the four benchmark datasets used and described in section \ref{s:comp_stud}: \texttt{PDAC} in top left, \texttt{BlREAL2} in top right, \texttt{BlMIX2016} in bottom left and \texttt{BlMIX2018} in bottom right.}
\end{figure*}

Figure \ref{f:true_beta} displays a heatmap of true proportions for all four benchmark datasets used and described in Section \ref{s:comp_stud}. Cell types and bulks are both reordered so that similar bulks in terms of cellular composition are grouped in clusters. 

The plot shows the two subgroup structure for both BlMIX datasets and for BlREAL2. For PDAC, it shows that fibroblasts are obviously dominant in all bulks and that they can be divided into two nearly balanced clusters, one with a much larger proportion of classical than basal cancer cells and the other one with more basal than classical cancer cells. 

\begin{figure*}[!ht]
  \begin{center}
  \includegraphics[width=\textwidth]{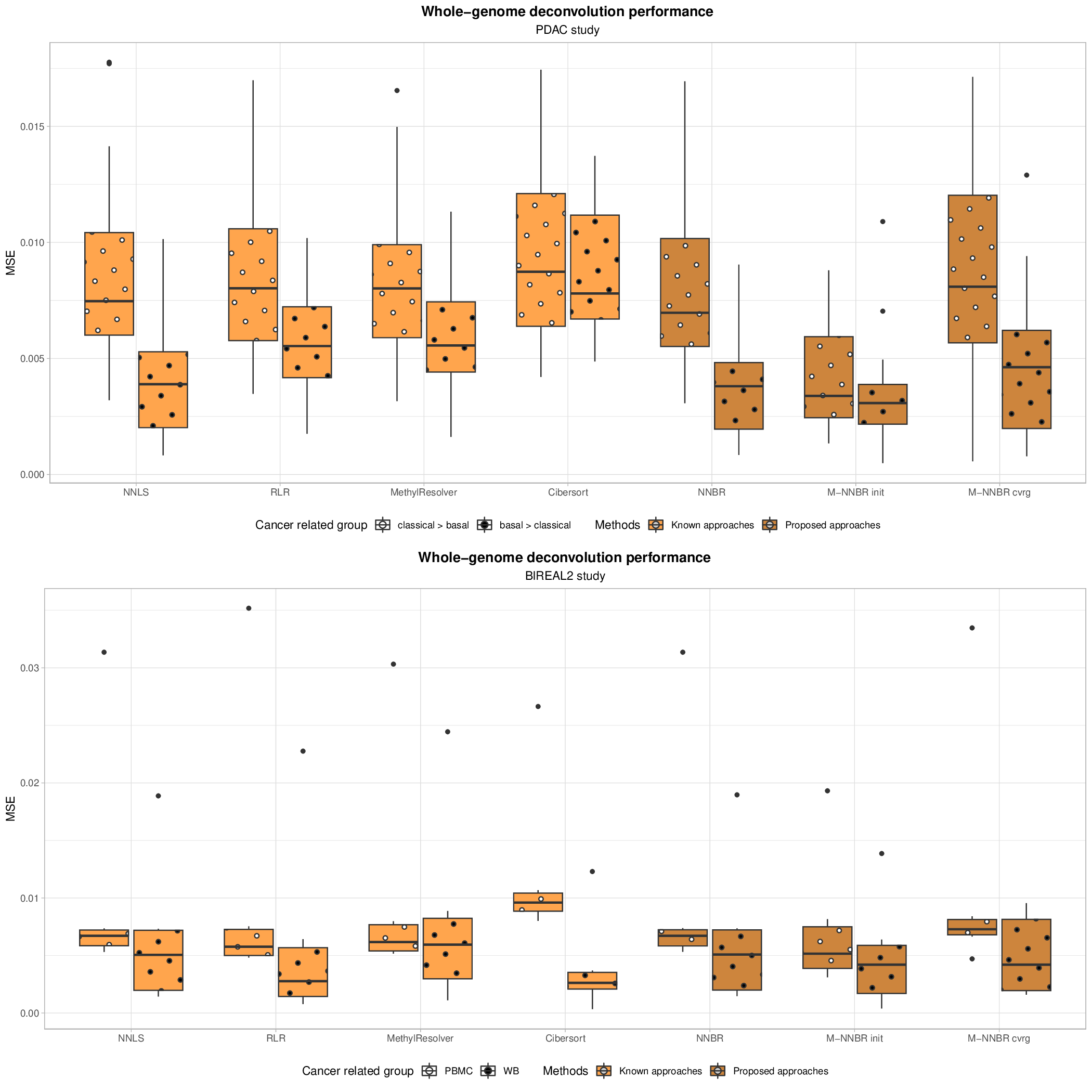}
  \end{center}
  \caption{\texttt{PDAC} and \texttt{BlREAL2} studies - Absolut mean Squared Error (MSE) for the respectively 30 and 12 bulks of selected cell deconvolution methods and the proposed mixture of non-negative Beta regression.}
  \label{fig:absolut_mse_by_subgroup}
\end{figure*}

\begin{figure*}[!ht]
  \begin{center}
  \includegraphics[width=\textwidth]{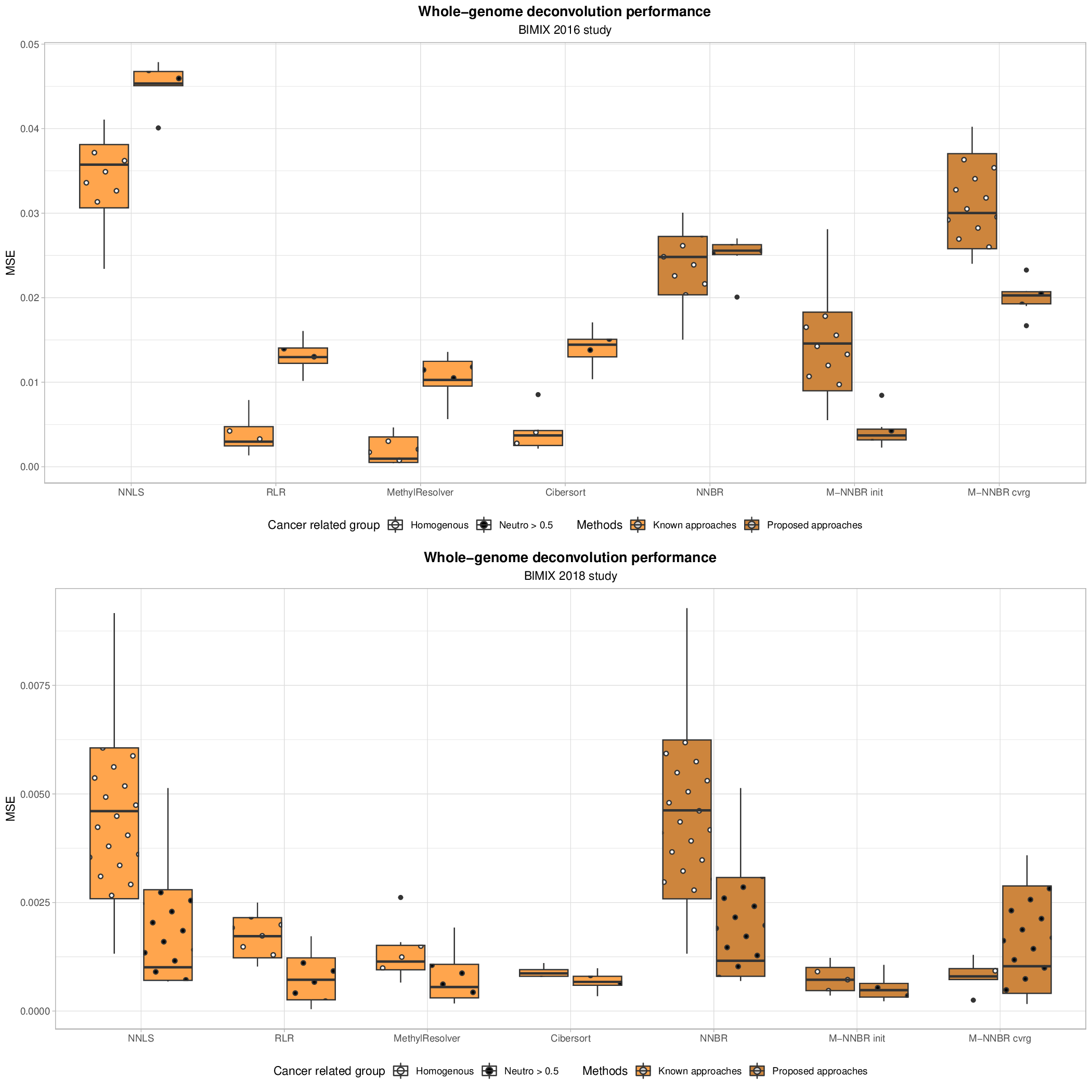}
  \end{center}
  \caption{\texttt{BlMIX2016} and \texttt{BlMIX2018} studies - Absolut mean Squared Error (MSE) for 12 bulks of selected cell deconvolution methods and the proposed mixture of non-negative Beta regression.}
  \label{fig:absolut_mse_by_group_whole_genome_vs_Mixt_NNBR2}
\end{figure*}

\section{Comparative studies on BlMIX 2016\label{app:BlMIX2016}}

\begin{figure*}[!ht]
  \begin{center}
  \includegraphics[width=\textwidth]{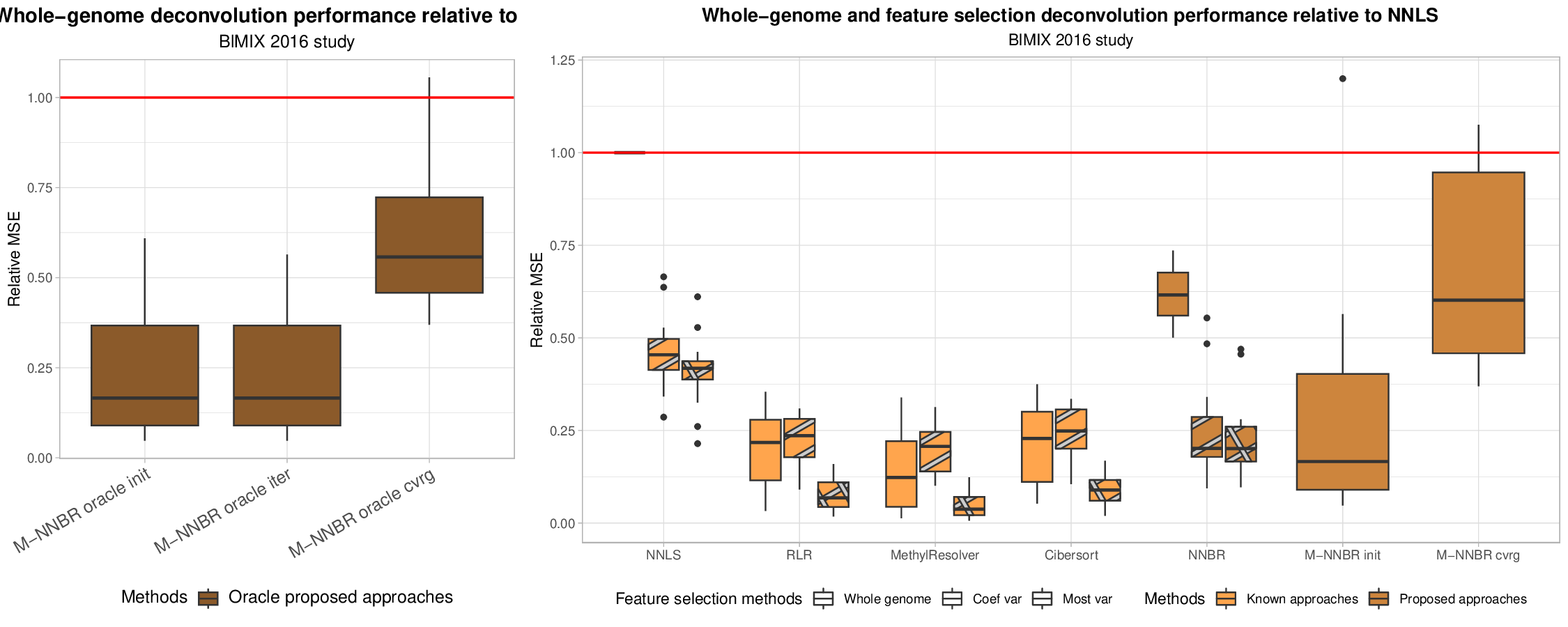}
  \end{center}
  \label{fig:mse_whole_genome_and_fs_vs_Mixt_NNBR_blmix2016}
  \caption{\texttt{BlMIX2016} study - Left plot: Relative efficiencies (RE) with respect to NNLS across all bulk samples. The component selected by the proposed mixture of non-negative Beta regression (M-NNBR) is chosen in an oracle manner for three types of partitions: at EM algorithm initialization, at the iteration yielding the best estimation accuracy, and at convergence. Right plot: Relative efficiencies (RE) of selected cell deconvolution methods with respect to NNLS across bulk samples. Two feature selection strategies, representative of commonly used statistical criteria in deconvolution pipelines, are evaluated for all known deconvolution approaches and NNBR: (i) selection based on the highest gene-wise coefficient of variation in the reference matrix, and (ii) selection based on the highest gene-wise variance.}
\end{figure*}

Comparative study of Section \ref{s:comp_stud} applied to \texttt{BlMIX2016} \textit{in vitro} dataset, related to \texttt{BlMIX2018} \textit{in vitro}  dataset. Figure \ref{fig:mse_whole_genome_and_fs_vs_Mixt_NNBR_blmix2016} displays on the left plot the relative efficiencies of the three oracle variant of M-NNBR and on the right plot the relative efficiencies comparison between all tested model with two feature selection strategies used on NNLS, RLR, MethylResolver, CIBERSORT and NNBR.

\end{document}